\documentclass[a4paper]{jpconf}
\usepackage{graphicx,cite,amsmath}

\newcommand{\afb}{A_{FB}}
\newcommand{\ac}{A_C}

\begin{document}
\title{$t \bar t$ charge asymmetry, family and friends\footnote{Talk given by J.A. Aguilar-Saavedra at Discrete 2012, Lisbon, Portugal, December 3-7 2012}}

\author{J A Aguilar-Saavedra, M P\'erez-Victoria}

\address{Departamento de F\'{\i}sica Te\'orica y del Cosmos, Universidad de Granada, 18071 Granada}

\ead{jaas@ugr.es,mpv@ugr.es}

\begin{abstract}
We present the current status of the Tevatron charge asymmetry and its sister asymmetry at the LHC. The relation between both is elucidated, using as framework the collider-independent asymmetries they originate from. Other related observables, such as the $t \bar t$ differential distribution and top polarisation, are also discussed.
\end{abstract}

\section{Introduction}
The measurement of an anomalous forward-backward (FB) asymmetry by the CDF and D0 experiments at the Tevatron~\cite{Aaltonen:2012it,Abazov:2011rq}
has fueled a plethora of proposals addressing the apparent deviation from the predictions of the Standard Model (SM). In most cases, these proposals involve new physics beyond the SM. But the simplest explanations for these departures, including those invoking higher-order SM corrections, also predict an enhancement of the charge asymmetry at the Large Hadron Collider (LHC). Such an enhancement has not been observed, though it is not excluded either, given the present experimental uncertainties. To make the situation even more puzzling, new physics models addressing the Tevatron excess often, but not always, predict related effects in $t \bar t$ production, such as enhancements in the high $t \bar t$ invariant mass tail, or a net polarisation of the top (anti)quark. None of these effects have been observed. In the following, we critically review these issues and discuss the intriguing status of the subject.

\section{The charge asymmetry at the Tevatron}

\begin{figure}[t]
\begin{center}
\begin{tabular}{ccc}
\includegraphics[height=6cm]{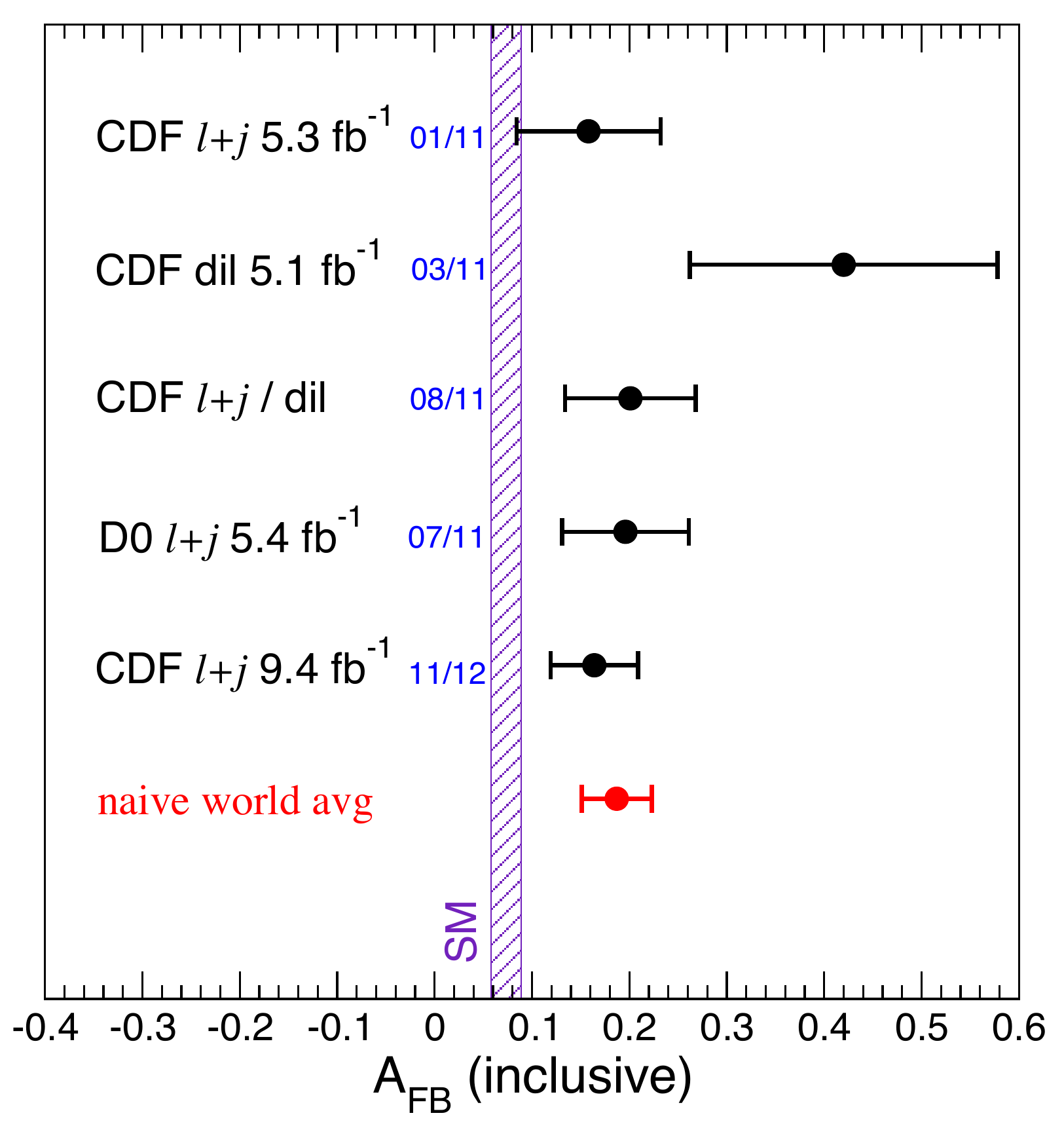} & \quad &
\includegraphics[height=6cm]{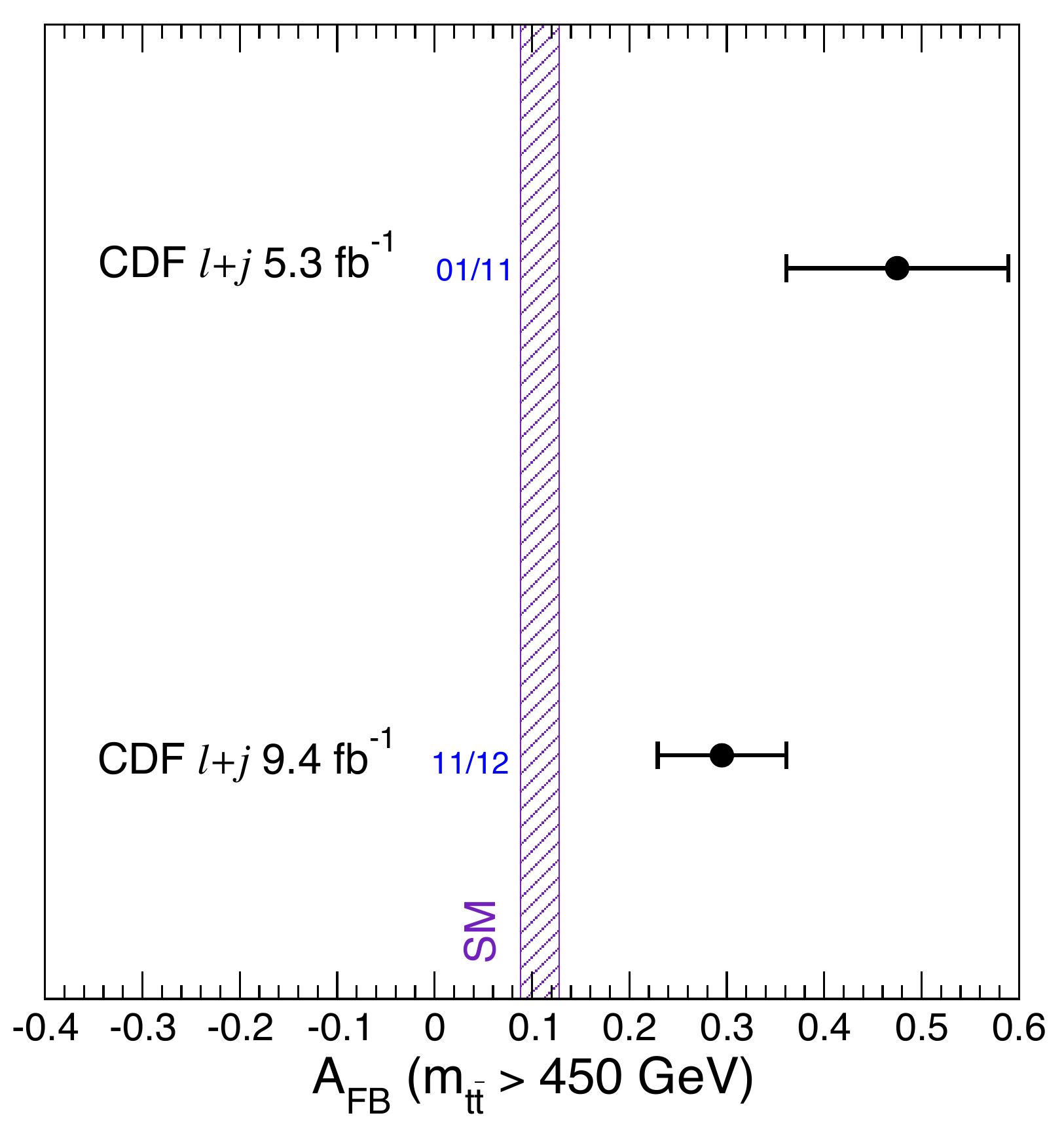}
\end{tabular}
\end{center}
\caption{\label{fig:tevX}Measurements of $\afb$ at the Tevatron: inclusive (left) and at high $m_{t\bar t}$ (right).}
\end{figure}

It is well known that the differential cross section for $q \bar q \to t \bar t$, with $q=u,d$, is not invariant under the interchange of the $t$ and $\bar t$ momenta. At the Tevatron, the most commonly used observable to measure this difference is the asymmetry
\begin{equation}
\afb = \frac{N(\Delta y>0) - N(\Delta y<0)}{N(\Delta y>0) + N(\Delta y<0)} \,,
\label{ec:afb}
\end{equation}
where $\Delta y = y_t - y_{\bar t}$ is the difference between the rapidities of the top quark and antiquark, taking the $z$ axis in the proton direction. This rapidity difference is invariant under boosts in the beam direction. The  definition above exploits the fact that, to a good extent, in $p\bar p$ collisions the directions of the initial quark and antiquark are known: they are, respectively, the directions of the proton and the antiproton. This asymmetry is equivalent to an asymmetry in the polar angle $\theta$ between the top momentum and the quark direction in the centre of mass (CM) frame,
\begin{equation}
\afb = \frac{N(\cos \theta >0) - N(\cos \theta<0)}{N(\cos \theta>0) + N(\cos \theta<0)} \,.
\label{ec:afb2}
\end{equation}
In the SM this asymmetry is small and arises, at the lowest order in perturbation theory, from the interference of tree-level and one-loop diagrams for $q \bar q \to t \bar t$, plus some contributions from extra jet radiation. For some time, the two Tevatron experiments have consistently measured an asymmetry above the SM expectation, in the semileptonic and dilepton $t \bar t$ decay channel. A summary of the most recent inclusive unfolded measurements is shown in Fig.~\ref{fig:tevX} (left), to be compared with the SM predictions~\cite{Campbell:2012uf,Ahrens:2011uf,Hollik:2011ps,Kuhn:2011ri,Bernreuther:2012sx}, which range from 0.058 to 0.089. The naive world average of the latest measurements in each experiment and $t \bar t$ decay channel gives an asymmetry of $0.187 \pm 0.036$, which is $2.7\sigma$ above the closest of those SM predictions.
In contrast with other recent anomalies, the trend of the measurements with time and increased luminosity does not approach the SM prediction but the positive excess persists and it does not look like a statistical fluctuation. At high $t \bar t$ invariant mass $m_{t \bar t} \geq 450$ GeV (Fig.~\ref{fig:tevX} right) the departure from the SM has been reduced with respect to the first measurement~\cite{Aaltonen:2011kc}, but the deviation is still significant. 

These consistent discrepancies have motivated a number of papers proposing new physics explanations. As the excess in $\afb$ is of the same size as the one-loop QCD asymmetry, if this excess results from new physics it is likely to enter at the tree level in $q \bar q \to t \bar t$. The types of renormalisable new physics that can enter $q \bar q \to t \bar t$ can be classified by using group theory, requiring that the Lagrangian is invariant under $\text{SU}(3)_C \times \text{SU}(2)_L \times \text{U}(1)_Y$. This requirement gives a total of 18 possibilities, 10 for spin-1 vector bosons and 8 for scalars~\cite{AguilarSaavedra:2011vw}. Among them, the most popular ones are a colour octet $G_\mu$ (also called ``gluon'' hereafter) exchanged in the $s$ channel~\cite{Djouadi:2009nb}, a $Z'$~\cite{Jung:2009jz}, $W'$~\cite{Cheung:2009ch} or scalar doublet $\phi$ ~\cite{Nelson:2011us} in the $t$ channel, and a colour sextet $\Omega^4$ or colour triplet scalar $\omega^4$~\cite{Shu:2009xf} in the $u$ channel. 
These ``simple'' models are phenomenological in the sense that their goal is to explain the anomaly simply by adding to the SM a new particle.\footnote{In addition to these simple models, proposals have been made with new physics entering at one loop level~\cite{Gabrielli:2011jf} that give similar predictions.} Still, they provide a good basis to test if:
\begin{enumerate}
\item $\afb$ can be enhanced without spoiling the good agreement of the total $t \bar t$ cross section with the SM prediction.
\item The inclusive and high-mass measurements of $\afb$ can be reproduced.
\item The explanation of the Tevatron anomalies is compatible with other data, in particular from the LHC.
\end{enumerate}
Once we have a model that fulfills these conditions, one can go further and build a more complete new physics model that explains the Tevatron measurements of the asymmetry and other related ones. We now concentrate on the first two tests, which involve Tevatron collider.

The first test that these models have to pass concerns their ability to correctly fit the observed total cross section and asymmetry. This is nontrivial since the SM prediction of the cross section, $\sigma_\text{SM} = 7.5 \pm 0.5$ pb~\cite{Aliev:2010zk} is very close to the measured one, $\sigma_\text{exp} = 7.68 \pm 0.4$~\cite{ttxsexp}. If one expands the cross section in the presence of new physics as
$\sigma = \sigma_\text{SM} + \delta\sigma_\text{int} + \delta \sigma_\text{quad}$, this implies that the interference between SM and new physics, $\delta\sigma_\text{int}$, and the new physics quadratic term, $\delta \sigma_\text{quad}$, have to fulfill $ \delta\sigma_\text{int} + \delta \sigma_\text{quad} \simeq 0$. There are two possibilities for this:
\begin{enumerate}
\item The quadratic term is large, which implies that the interference is also large and both terms nearly cancel each other. For this to happen, a fine-tuning of the new physics coupling is needed, since the interference and quadratic terms depend linearly and quadratically, respectively, on this coupling. This is the case, for example, of $t$-channel models ($Z'$, $W'$, $\phi$). If this cancellation is arranged to happen at the Tevatron energy, then it is not expected to take place anymore at the LHC.
\item The quadratic term is small. The total interference $\delta \sigma_\text{int}$ must also be small but with  sizeable contributions in the forward and backward hemispheres, $\delta \sigma_\text{int}^F \simeq - \delta \sigma_\text{int}^B$. This is the case, for example, of an $s$-channel heavy gluon with axial coupling to either light quarks or to the top quark, in which case the interference identically vanishes. If one drops the condition $\delta \sigma_\text{int}^F + \delta \sigma_\text{int}^B \simeq 0$, as for example in $u$-channel colour sextet models, the generated asymmetries have to be smaller due to the total cross section constraint.
\end{enumerate}
In both cases, it is evident that one needs that the interference with the SM is non-zero~\cite{Grinstein:2011yv} which, in principle, can be achieved with all the types of new particles that can contribute to $q \bar q \to t \bar t$ ~\cite{AguilarSaavedra:2011vw}.

\begin{figure}[htb]
\begin{center}
\begin{tabular}{ccc}
\includegraphics[height=4.8cm]{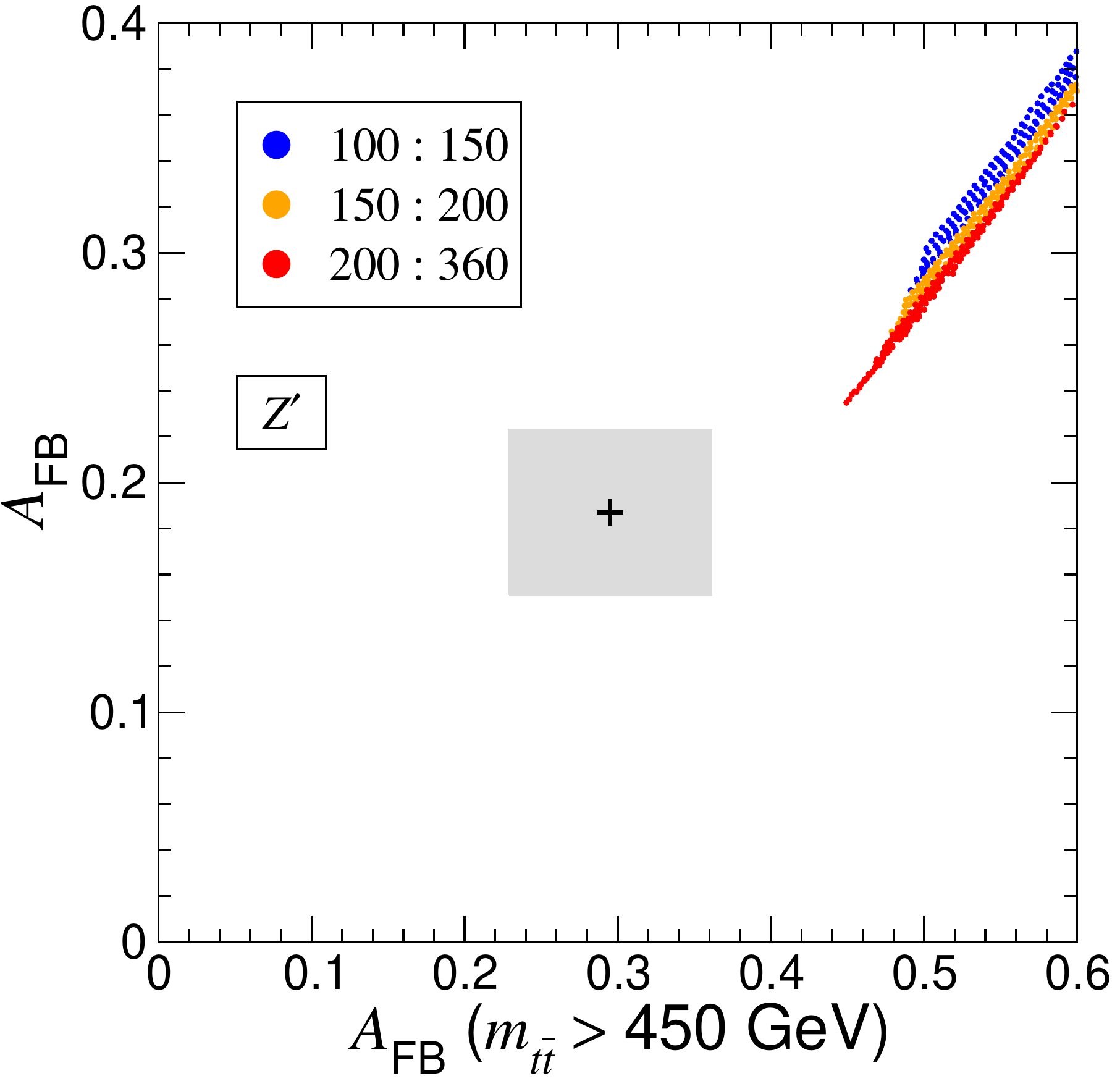} &
\includegraphics[height=4.8cm]{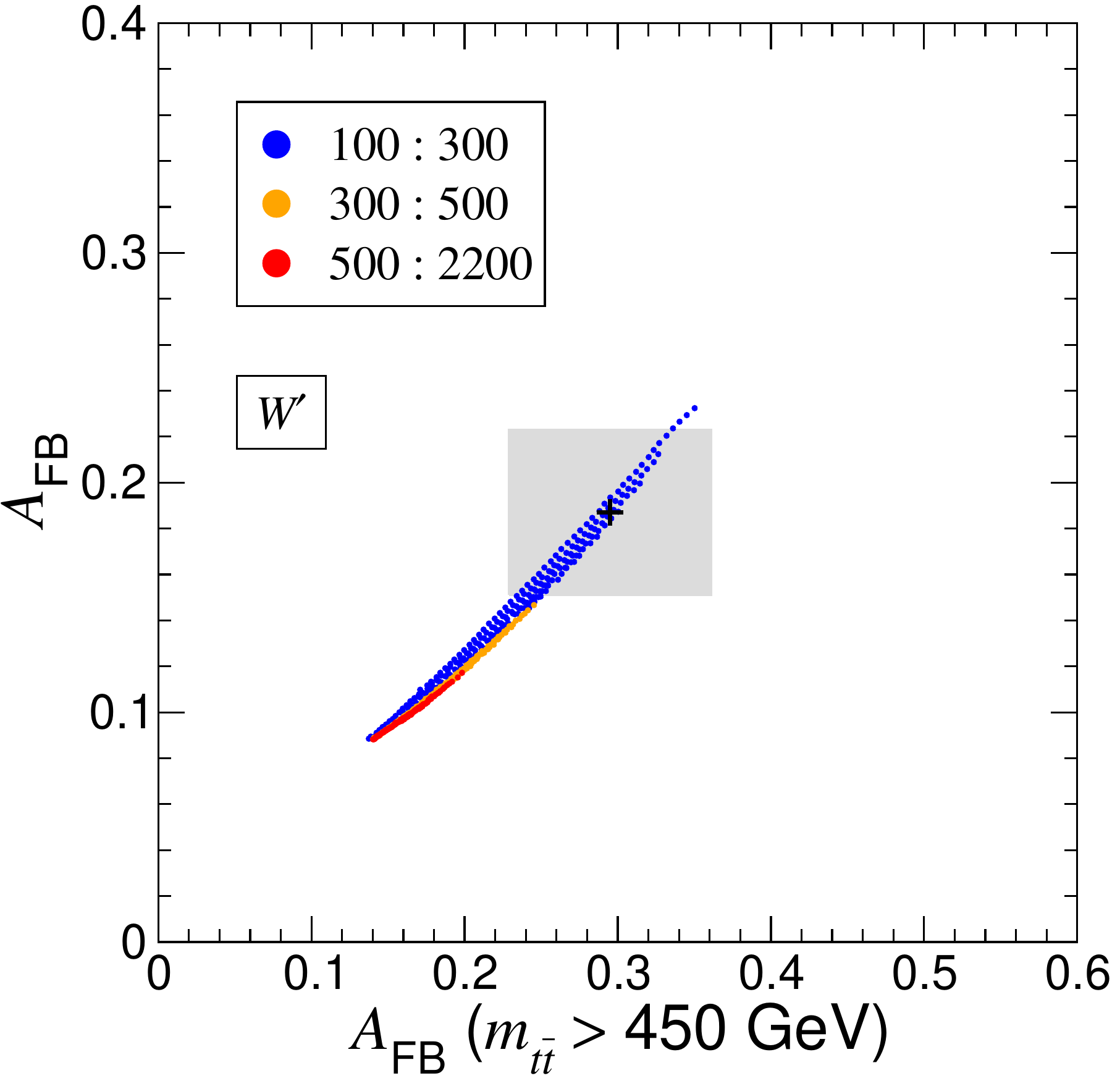} &
\includegraphics[height=4.8cm]{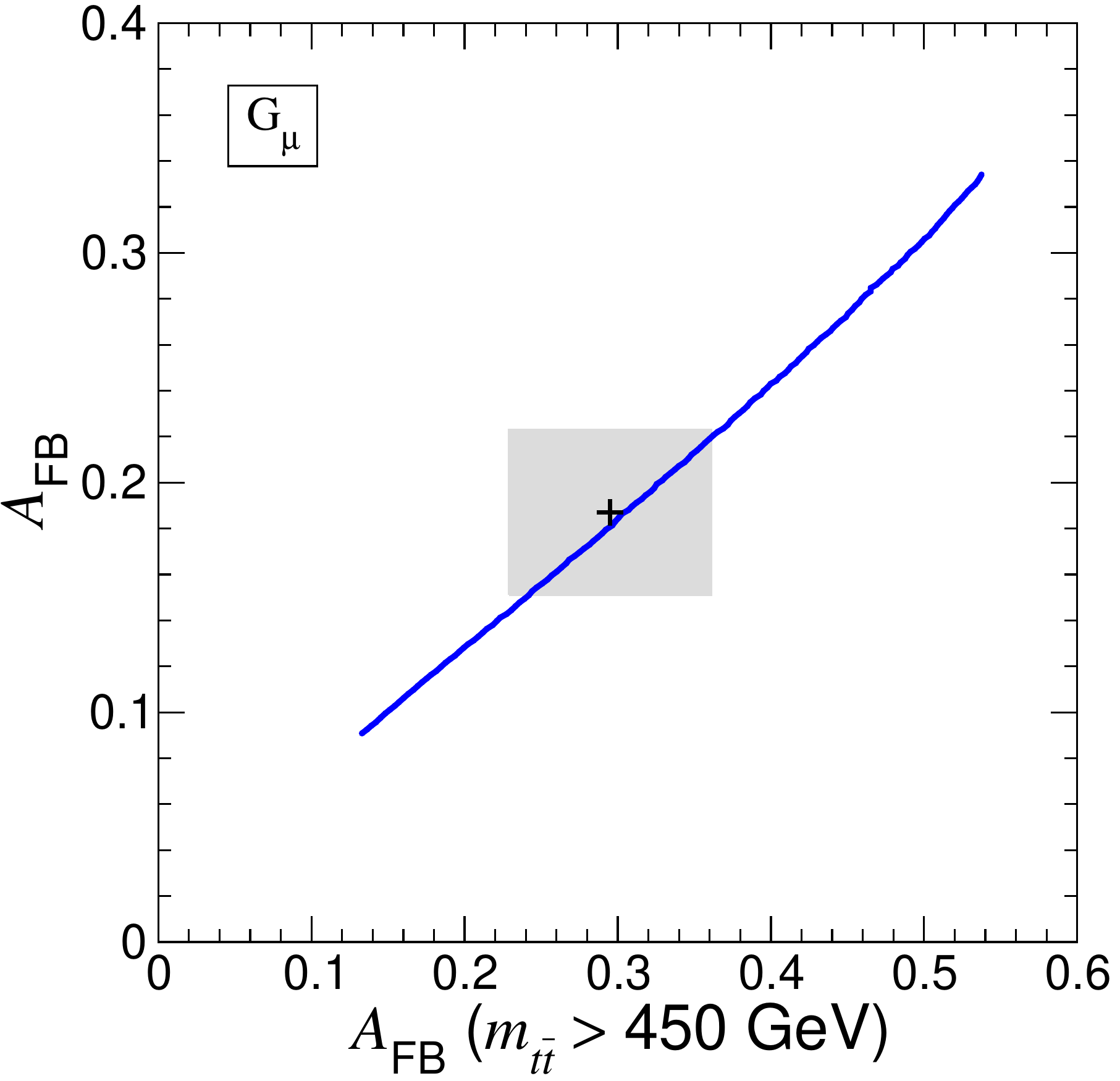} \\
\includegraphics[height=4.8cm]{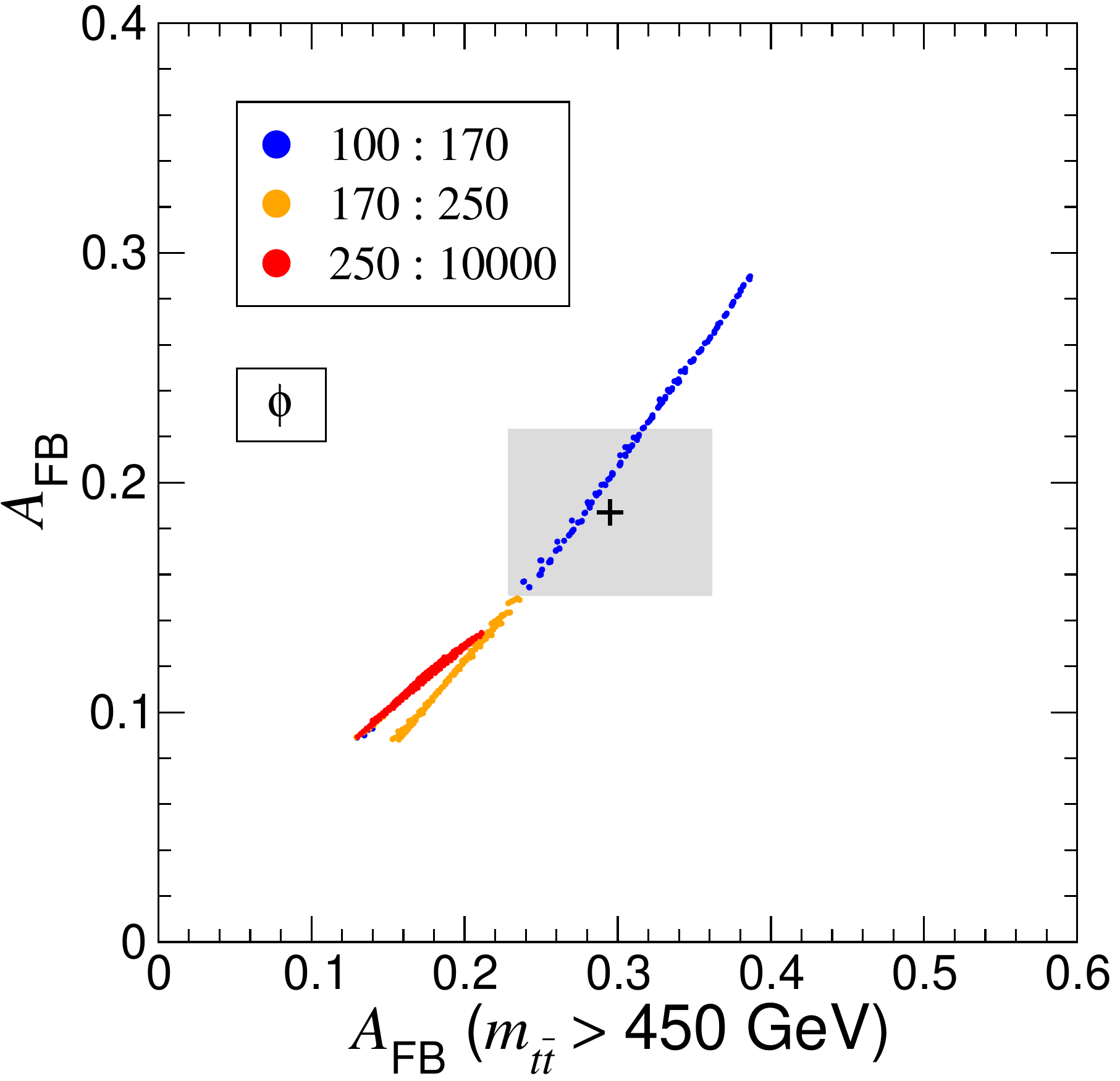} & 
\includegraphics[height=4.8cm]{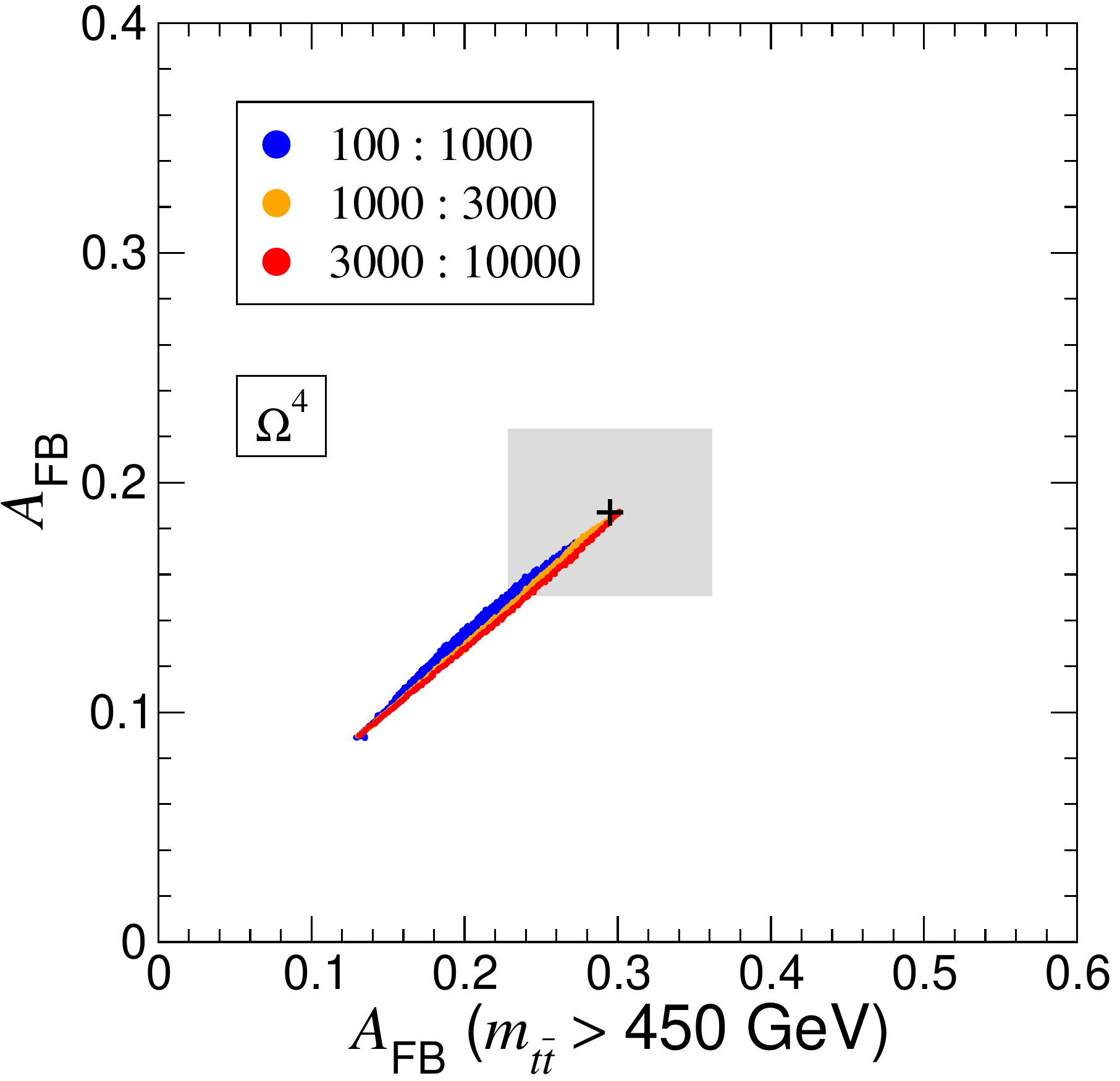} &
\includegraphics[height=4.8cm]{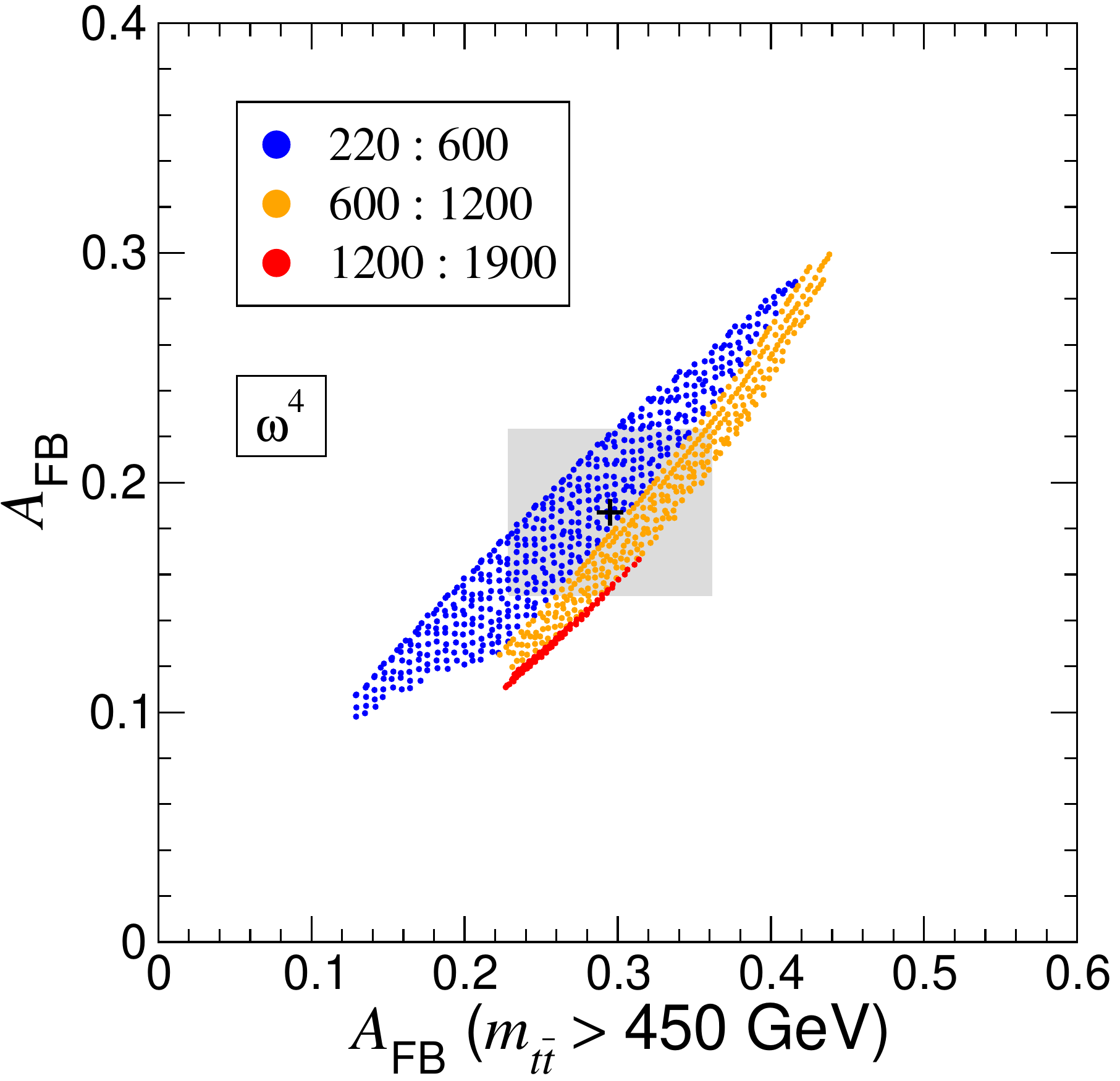}
\end{tabular}
\end{center}
\caption{\label{fig:AvsAH}Inclusive versus high-mass asymmetries at the Tevatron, for several new physics models. The numbers in the legends indicate the mass range for the new particle, in GeV. The crosses correspond to the experimental measurements in Fig.~\ref{fig:tevX}, with the shaded boxes corresponding to the $1\sigma$ uncertainty.}
\end{figure}

A second Tevatron test for the new physics proposals concerns whether they can accommodate the inclusive and high-mass measurements. This is accomplished for most models, as it is shown in Fig.~\ref{fig:AvsAH}. The coloured regions show the model predictions obtained by a parameter space scan, subject to some loose constraints from the total $t \bar t$ cross section at the Tevatron and the high-mass tail at the LHC~\cite{AguilarSaavedra:2011ug}. We only consider positive contributions to the asymmetry, which are summed to the SM one~\cite{Bernreuther:2012sx} in all cases. For the colour octet we consider a very heavy axigluon represented by four-fermion operators~\cite{AguilarSaavedra:2010zi,Degrande:2010kt,AguilarSaavedra:2011vw,AguilarSaavedra:2011zy}.
Most of these simple models can reproduce very well the inclusive and high-mass asymmetries, except a new $Z'$ which overpredicts the asymmetry, especially at high $t \bar t$ invariant mass mass. (For this and other reasons ---see section~\ref{sec:tail}--- this model will not be considered further here as a viable candidate.) But a more revealing outcome of this comparison is the observation that the inclusive and high-mass measurements of the asymmetry are ``naturally consistent'' or, in other words, it does not take a contrived model to reproduce both.\footnote{More complicated models,  for example with a number of $s$-channel coloured resonances, can reproduce complicated profiles of $\afb$ versus $m_{t \bar t}$~\cite{AguilarSaavedra:2011ci}.} A recent analysis of the polar angle dependence of the cross section~\cite{CDF-theta} also supports the internal consistence of the deviation. The differential cross section can be expanded in terms of Legendre polynomials,
\begin{equation}
\frac{d\sigma}{d\cos \theta} = \sum_l a_l P_l(\cos \theta) \,.
\end{equation}
The CDF Collaboration finds good agreement of all $a_l$ with the SM prediction, except for the term with $l=1$ ---corresponding to a term linear in $\cos \theta$--- that deviates more than $2\sigma$ from the SM. This pattern can be nicely fitted with an $s$-channel colour octet, for example, which enhances $a_1$ while keeping higher $l$ coefficients small.

Finally, at the Tevatron there are additional constraints from the cross section at the high-$m_{t\bar t}$ tail, but the measurements in that region are not very precise and the new physics contributions may have a much lower detection efficiency than SM top pair production, due to the limited detector acceptance~\cite{Gresham:2011pa}. This is the case, for example, of light $t$-channel mediators. For this reason we do not use the measurements in that region as a further constraint.

\section{The younger sister: the LHC charge asymmetry}

\begin{figure}[t]
\begin{minipage}{7.5cm}
\begin{center}
\includegraphics[height=6.3cm]{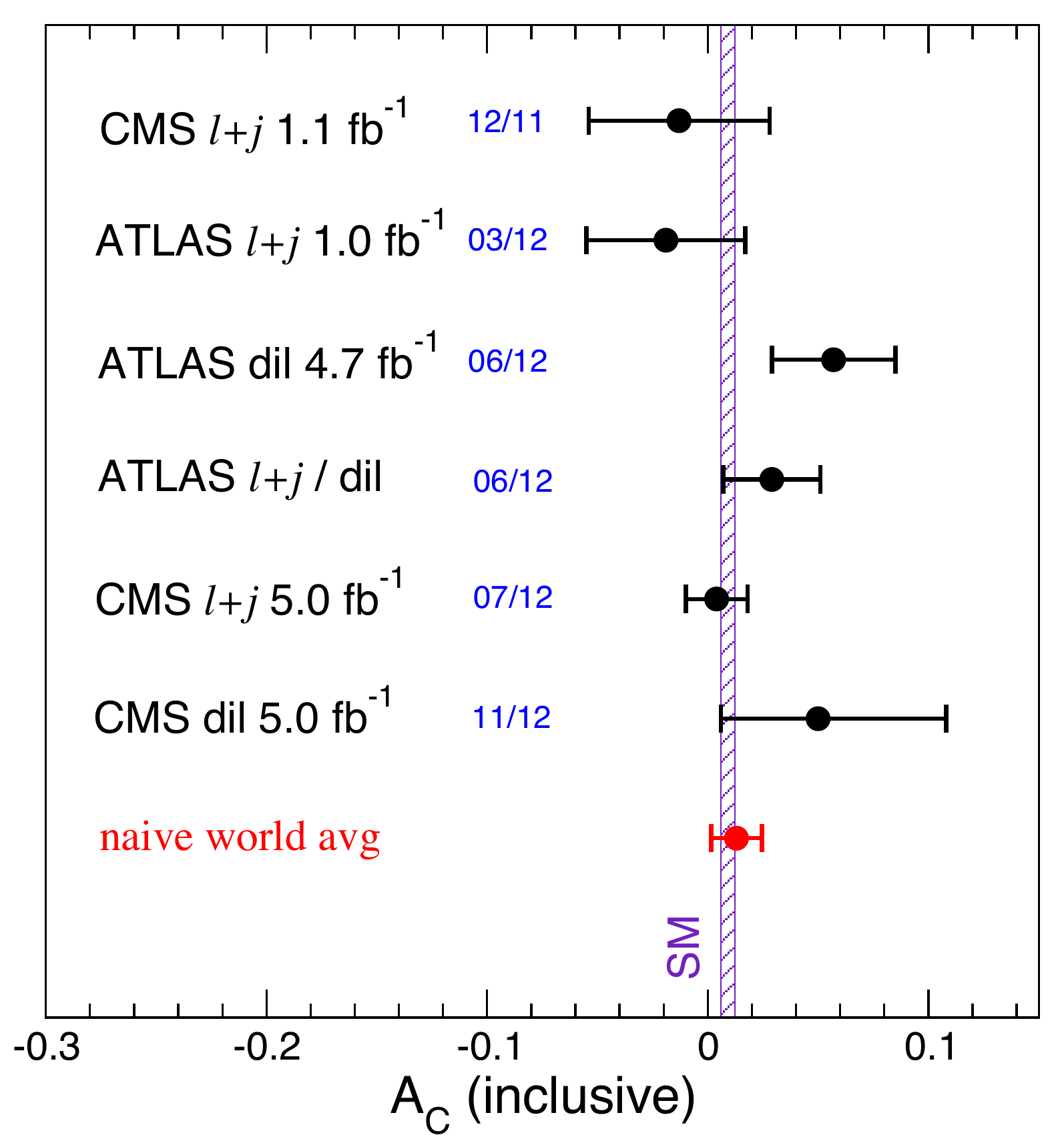}
\caption{\label{fig:acX}Measurements of $\ac$ (inclusive) at the LHC.}
\end{center}
\end{minipage}\hspace{2pc}%
\begin{minipage}{7.5cm}
\includegraphics[height=6cm]{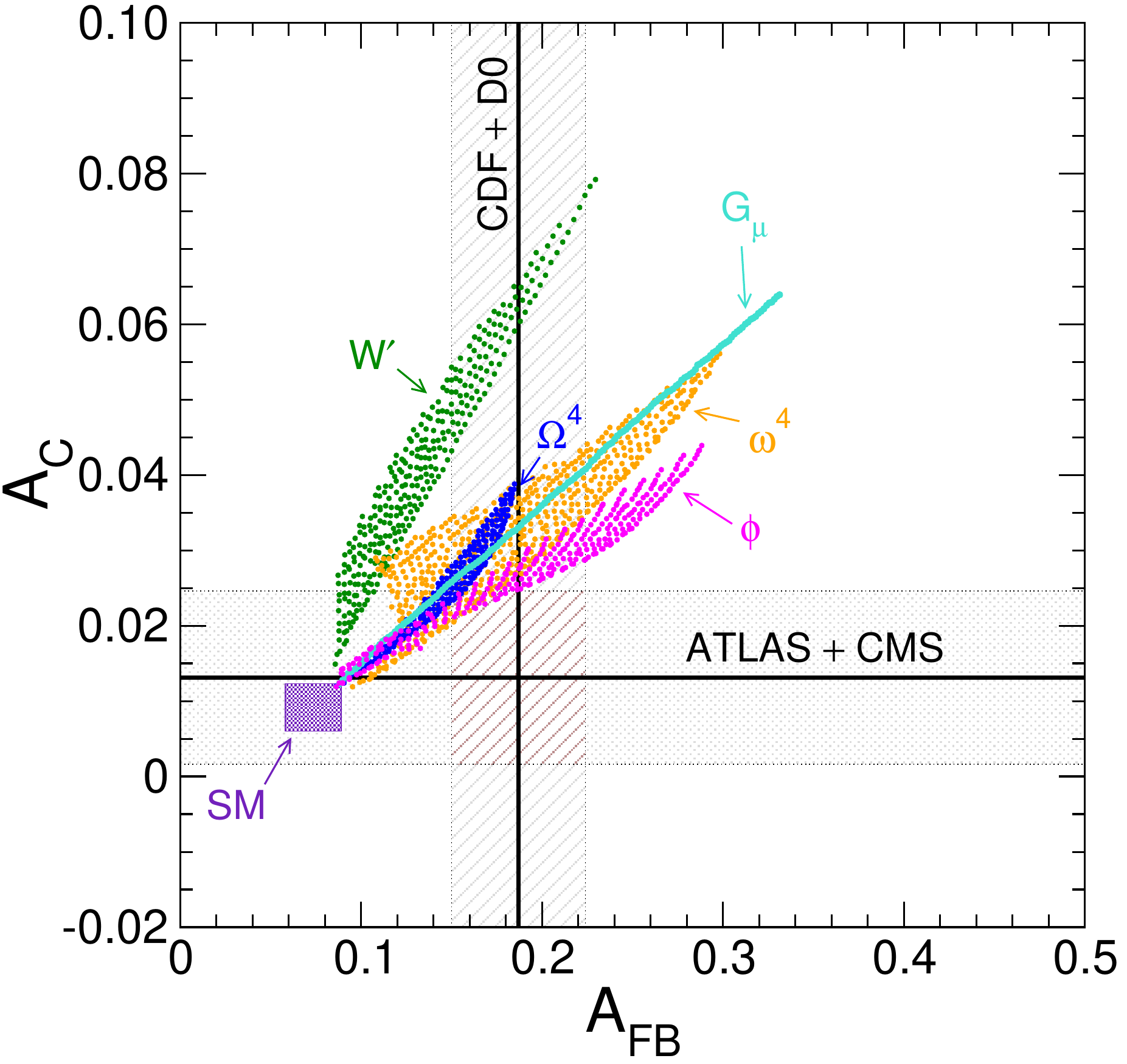}
\caption{\label{fig:AFBvsAC}Comparison of predictions for the inclusive asymmetries $\afb$ and $\ac$ for several simple models.}
\end{minipage} 
\end{figure}

At a $pp$ collider as the LHC, the symmetry of the initial state implies that, for a fixed choice of $z$ axis to measure rapidities, the FB asymmetry in Eq.~(\ref{ec:afb}) vanishes. Then, one has to consider different observables to test an asymmetry in $q \bar q \to t \bar t$. This can be done, for example, by exploiting the fact that valence quarks have larger average momentum fraction than antiquarks, leading to a non-vanishing asymmetry~\cite{Diener:2009ee}
\begin{equation}
\afb = \frac{N(\Delta |y|>0) - N(\Delta |y|<0)}{N(\Delta |y|>0) + N(\Delta |y|<0)} \,,
\label{ec:ac}
\end{equation}
with $\Delta |y| = |y_t| - |y_{\bar t}|$. This asymmetry has been measured by the ATLAS and CMS Collaborations, with the results shown in Fig.~\ref{fig:acX}. The naive average of the latest measurements is $\ac = 0.013 \pm 0.011$, consistent with the SM predictions~~\cite{Frixione:2002ik,Kuhn:2011ri,Bernreuther:2012sx}.

It is important to stress here a fact that should be obvious: the Tevatron $\afb$ and the LHC $\ac$ are {\it not} the same observable. Therefore, a measurement of $\ac$ consistent with the SM is not in conflict with a Tevatron excess~\cite{AguilarSaavedra:2012va}. On the other hand, comparing predictions for $\afb$ and $\ac$ {\it within a given model} brings important consequences for the model~\cite{AguilarSaavedra:2011hz}, as it is clearly depicted in Fig.~\ref{fig:AFBvsAC}.
In particular, the $W'$ models are clearly disfavoured, as they predict values of $\ac$ more than $3\sigma$ above present data when accommodating the Tevatron asymmetry. For the rest of simple models the fate is uncertain as they are consistent with data at the $2\sigma$ level. The difficulty to simultaneously reproduce the central values of $\afb$ and $\ac$ has  motivated the appearance of several less simple
 models~\cite{Ko:2012ud,Drobnak:2012cz} that can accommodate the central values of the Tevatron and LHC asymmetries, by introducing some type of cancellation among contributions. (See also Refs.~\cite{Drobnak:2012rb,Alvarez:2012ca}.)

\section{The parents: the collider-independent asymmetries.}

\begin{figure}[htb]
\begin{center}
\begin{tabular}{ccc}
\includegraphics[height=5.2cm]{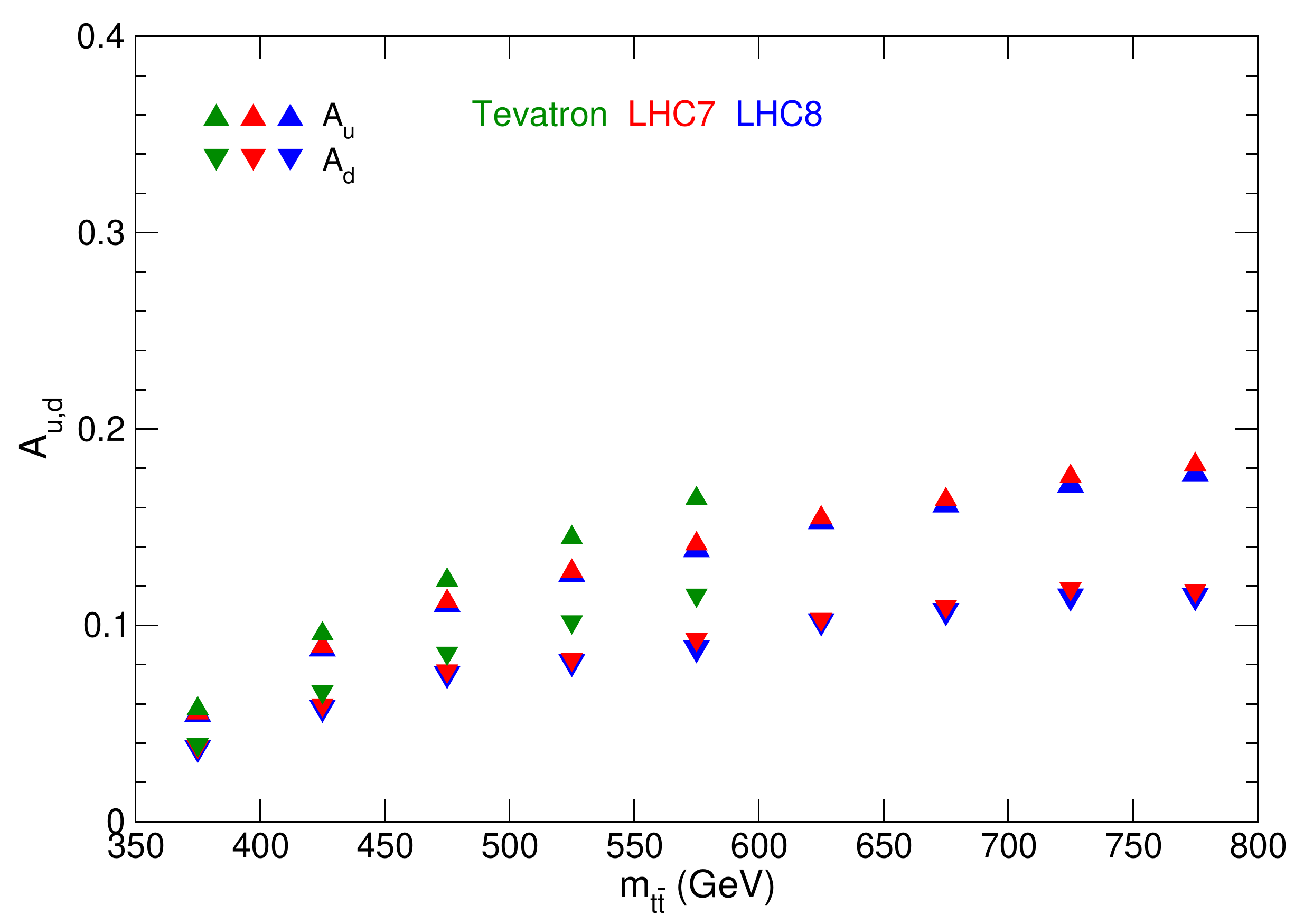} & \quad &
\includegraphics[height=5.2cm]{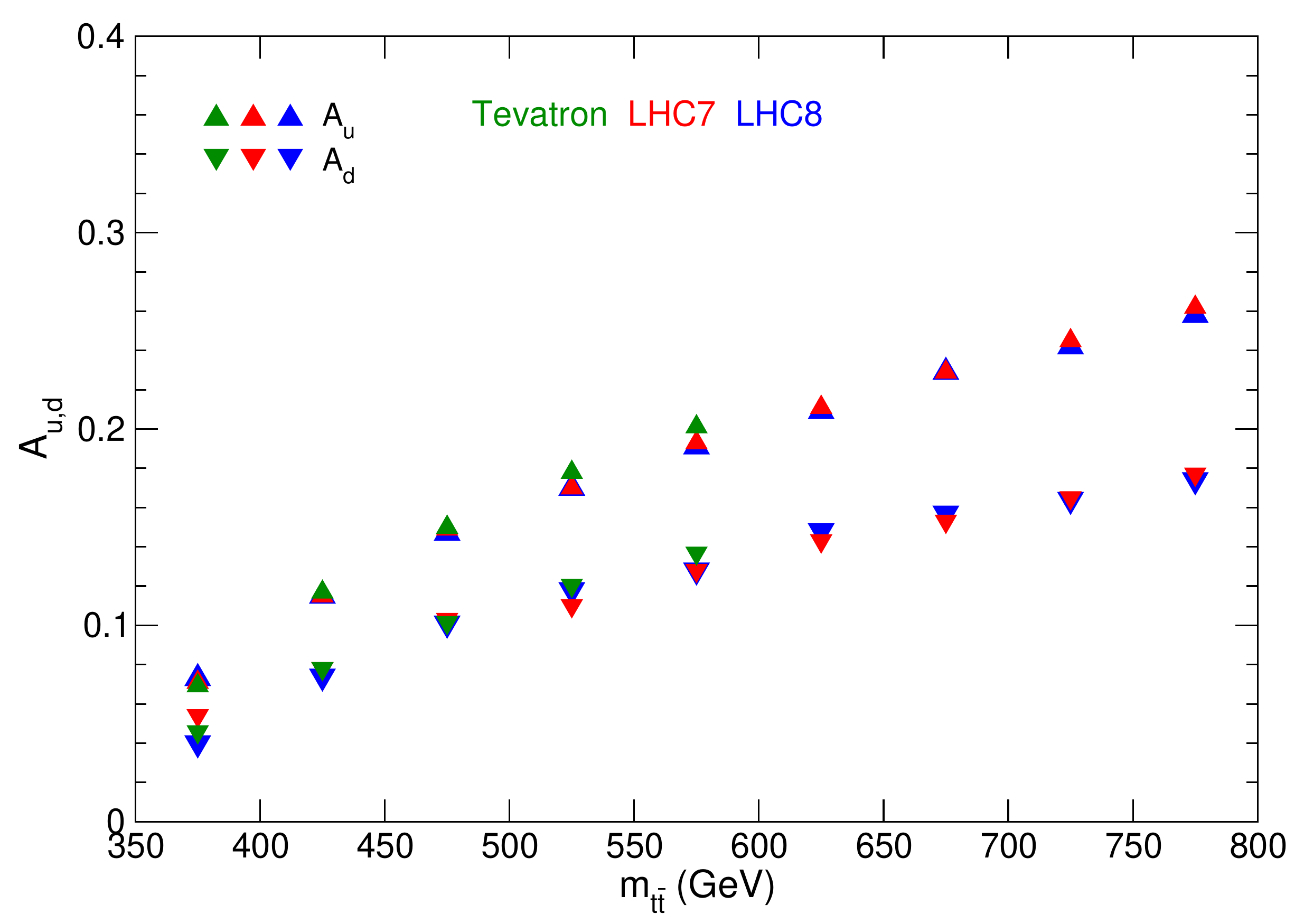}
\end{tabular}
\end{center}
\caption{\label{fig:AuAdSM}Asymmetries $A_u$, $A_d$ in the SM, for the Tevatron and the LHC. Left: without a cut on $p_T^{t \bar t}$. Right: for $p_T^{t \bar t} < 30$ GeV.}
\end{figure}

The Tevatron asymmetry $\afb$ and the LHC asymmetry $\ac$ originate from the ``intrinsic'' partonic asymmetries in $u \bar u \to t \bar t$, $d \bar d \to t \bar t$, which will be denoted hereafter as $A_u$, $A_d$, respectively~\cite{AguilarSaavedra:2012va}. At leading order (LO), these asymmetries only depend on the partonic CM energy $\hat s$. As a consequence of this, for a suitably narrow interval of $m_{t \bar t}$ the asymmetries $A_u$, $A_d$ are nearly the same at the Tevatron and the LHC. The ``daughter'' asymmetries $\afb$, $\ac$ can be regarded as different combinations of $A_u$ and $A_d$, the differences arising because
\begin{itemize}
\item at these two colliders the importance of $u \bar u \to t \bar t$ and $d \bar d \to t \bar t$, relative to the total $t \bar t$ production rate, changes due to parton density functions (PDFs);
\item at the LHC the asymmetry $\ac$ suffers from a ``dilution'' because not always the initial valence quark has larger momentum fraction than the sea antiquark. In case that the antiquark has larger momentum fraction, a ``forward'' event, that is, with the top quark in the direction of the incoming quark ($\cos \theta > 0$), has $\Delta |y|<0$ and contributes negatively to $\ac$.
\end{itemize}
Then, a possible experimental test of the consistency of $\afb$ (higher than the SM prediction) with $\ac$ (consistent with the SM) would be to measure the ``collider-independent'' asymmetries $A_u$ and $A_d$.\footnote{Although we use the name ``collider-independent'' for $A_u$ and $A_d$, to be precise it must be noted that at next-to-leading order (NLO) some differences are introduced, of little relevance from a practical point of view. These are mainly originated from the need to replace a fixed partonic CM energy $\hat s$ by a narrow $m_{t \bar t}$ interval, which introduces some deviations due to a residual dependence on PDFs. Besides, the asymmetries in $g q \to t \bar t j$ are irrelevant. For an extended discussion see Ref.~\cite{AguilarSaavedra:2012rx}.}
In experiments, the intrinsic asymmetries $A_u$, $A_d$ can be extracted by exploiting the dependence of $\afb$ and $\ac$ on the velocity of the $t \bar t$ pair~\cite{AguilarSaavedra:2011cp}
\begin{equation}
\beta = \frac{|p_t^z + p_{\bar t}^z|}{E_t + E_{\bar t}} \,,
\end{equation}
because $A_u$ and $A_d$ are independent of this variable.\footnote{Strictly speaking, they are independent at LO and with fixed $\hat s$. At NLO, or in finite $m_{t \bar t}$ bins, one can see that the dependence is rather mild~\cite{AguilarSaavedra:2012rx}.} Thus, $A_u$ and $A_d$ can be extracted from a fit to
\begin{eqnarray}
\afb(\beta) & = & A_u F_u (\beta) + A_d F_d (\beta) \,, \nonumber \\
\ac(\beta) & = & A_u F_u(\beta) D_u(\beta) + A_d F_d(\beta) D_d(\beta) \,,
\label{ec:AuAd}
\end{eqnarray}
where $F_{u,d}$ are the fractions of $u \bar u$ and $d \bar d$ events, respectively, and $D_{u,d}$ are factors to reflect the dilution of the asymmetries in $pp$ collisions we have mentioned above. Both $F_{u,d}$ and $D_{u,d}$ can be computed with a Monte Carlo within the SM and used as input to extract $A_u$ and $A_d$ from data. The SM predictions for these asymmetries are given in Fig.~\ref{fig:AuAdSM}. The left panel shows the asymmetries in 50 GeV bins, without any restriction on the transverse momentum of the $t \bar t$ pair $p_T^{t \bar t}$. It is important to remark here that, compared with their expected experimental uncertainty, the differences between the Tevatron and LHC asymmetries $A_u$, $A_d$ are irrelevent and justifiy labelling these asymmetries as ``collider-independent''. The right panel shows the asymmetries with a cut $p_T^{t \bar t} < 30$ GeV that practically eliminates the deviations between the Tevatron and LHC asymmetries.

\begin{figure}[t]
\begin{minipage}{7.5cm}
\begin{center}
\includegraphics[height=5cm]{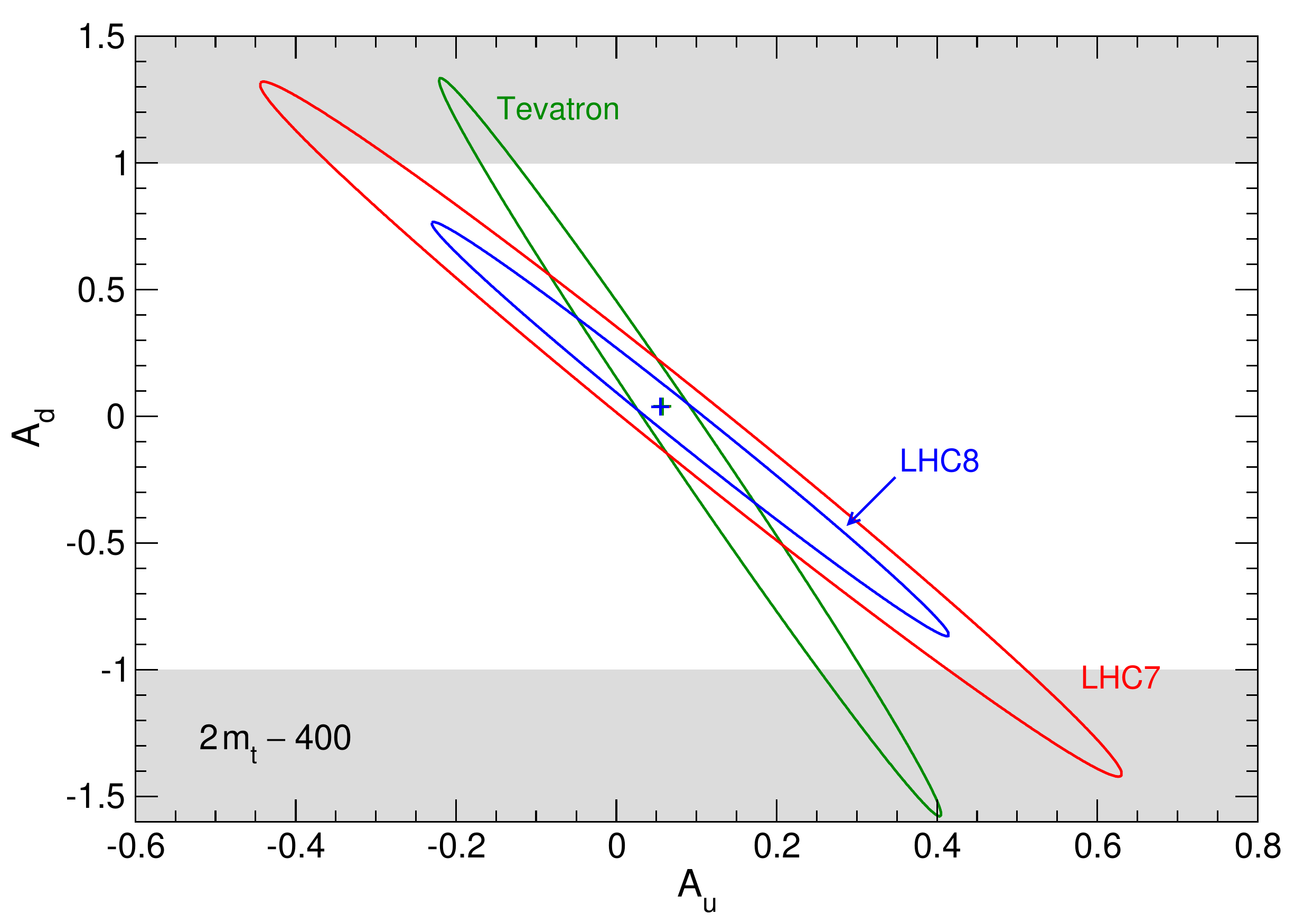}
\caption{\label{fig:AuAd1}Example of a potential outcome of the measurement of $A_u$ and $A_d$ at the Tevatron and the LHC.}
\end{center}
\end{minipage}\hspace{2pc}%
\begin{minipage}{7.5cm}
\begin{center}
\includegraphics[height=5cm]{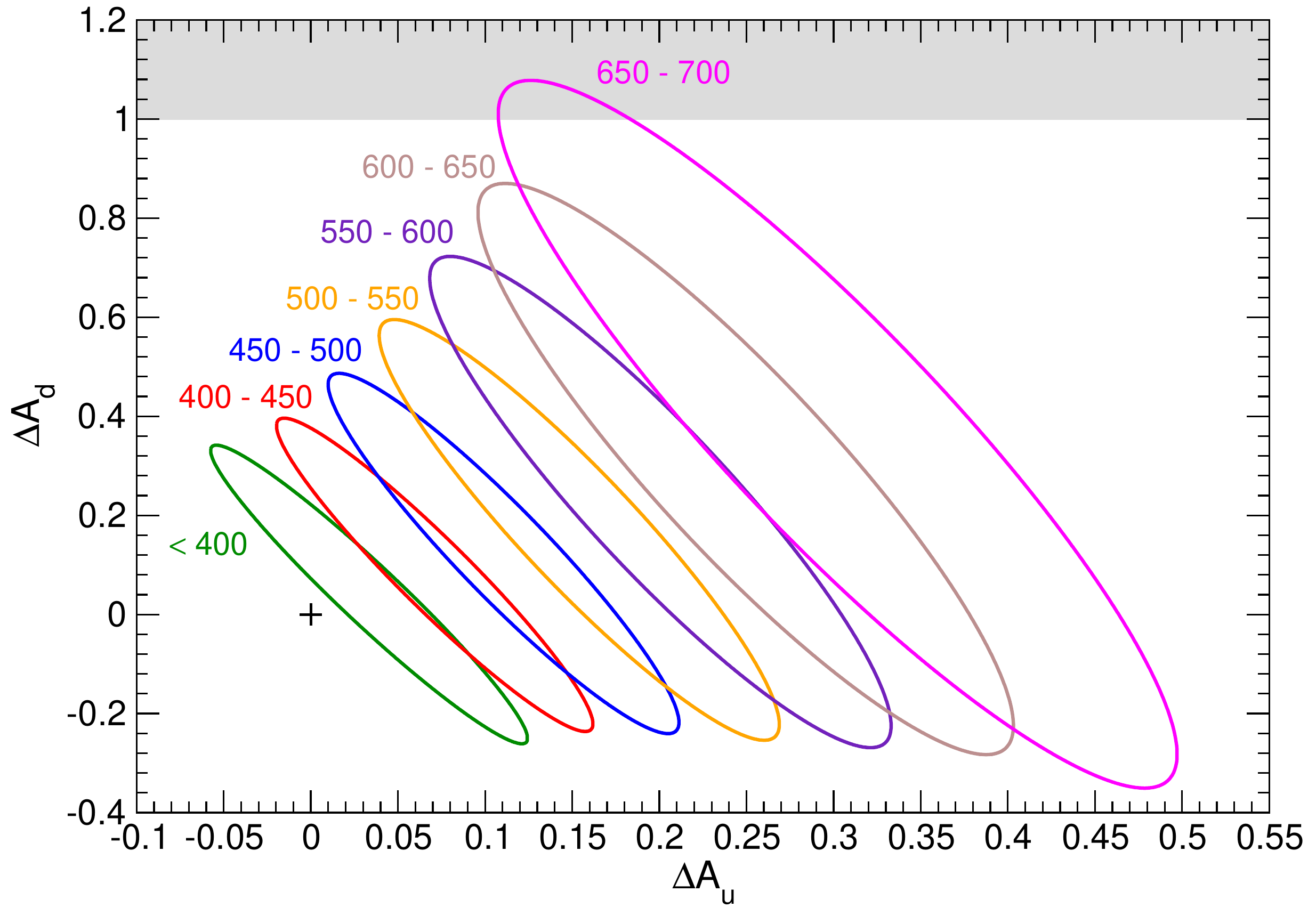}
\caption{\label{fig:Acomb}Example of a potential combination of the measurements of $A_u$ and $A_d$ at the Tevatron and the LHC.}
\end{center}
\end{minipage} 
\end{figure}

The experimental measurement of $A_u$ and $A_d$ is very challenging, as it may require a three-dimensional data unfolding in $\beta$, $m_{t \bar t}$ and $\Delta y$ ($\Delta |y|$). But the interest of this measurement may well be worth the effort. In the first place, if these asymmetries are measured, it can be tested whether the Tevatron and LHC results are consistent. We present in Fig.~\ref{fig:AuAd1} a potential result in the bin $m_{t \bar t} \leq 400$ GeV, with the crosses representing the NLO SM result and the ellipses corresponding to the $1\sigma$ statistical uncertainty. In this example the ellipses intersect by construction, since the SM NLO values of $\afb$ and $\ac$ have been used as input. But, if there is an unknown systematic effect in either collider, this may not be the case.
If, on the other hand, the Tevatron and LHC determinations of the asymmetries are consistent, one can combine both and test whether the combined measurement is compatible with the SM prediction. We show in Fig.~\ref{fig:Acomb} a potential result of a Tevatron-LHC combination, taking a heavy axigluon as a new physics benchmark. The ellipses correspond to the $1\sigma$ combined limits with the SM asymmetries subtracted, so that the SM point in this plot corresponds to zero. It is apparent that the combination of Tevatron and LHC results on $A_u$, $A_d$ has a much higher significance than the individual measurements.

\section{An old friend: the high-mass tail at the LHC} 
\label{sec:tail}

\begin{figure}[t]
\begin{center}
\begin{tabular}{ccc}
\includegraphics[height=4.8cm]{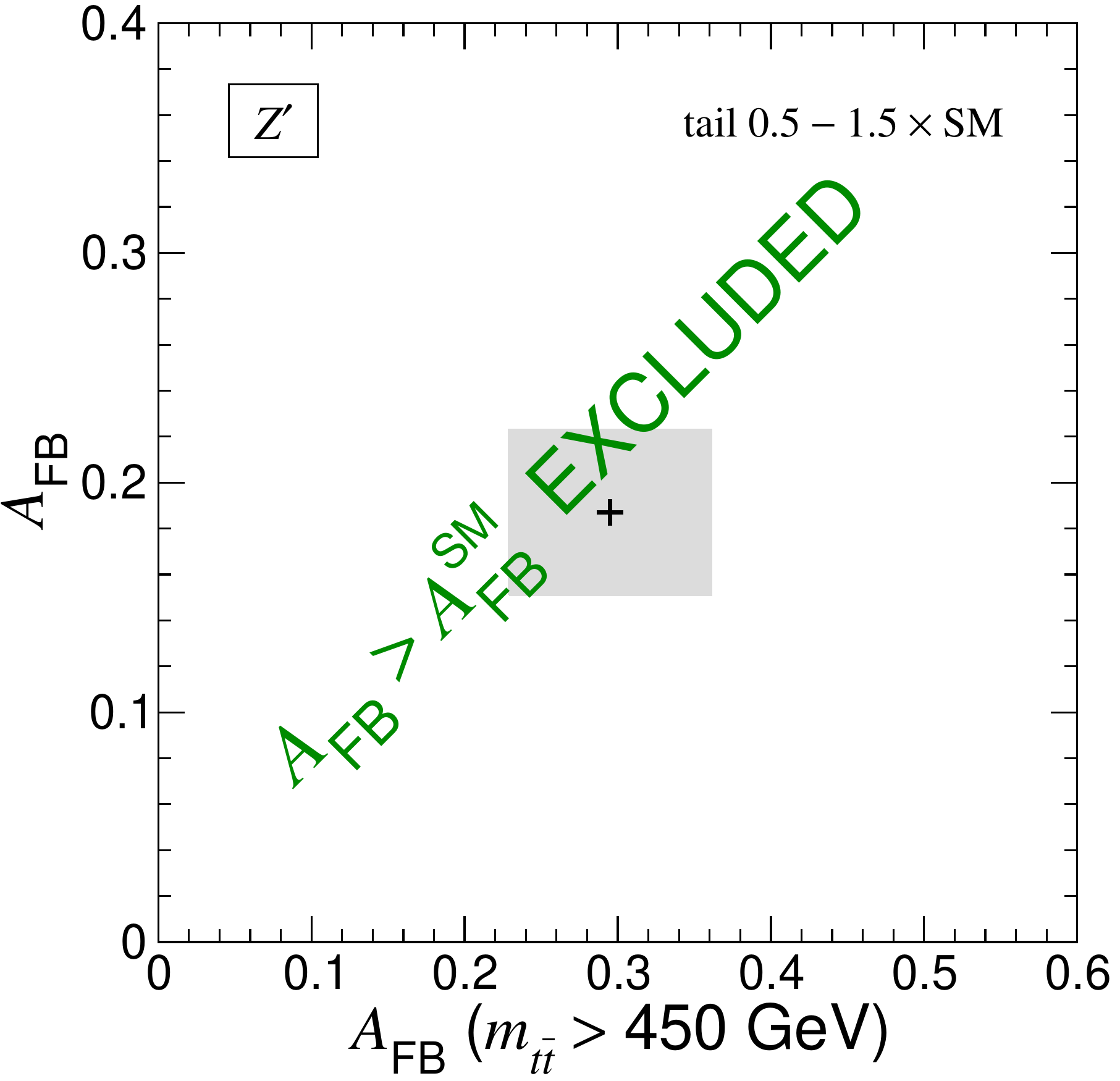} & \includegraphics[height=4.8cm]{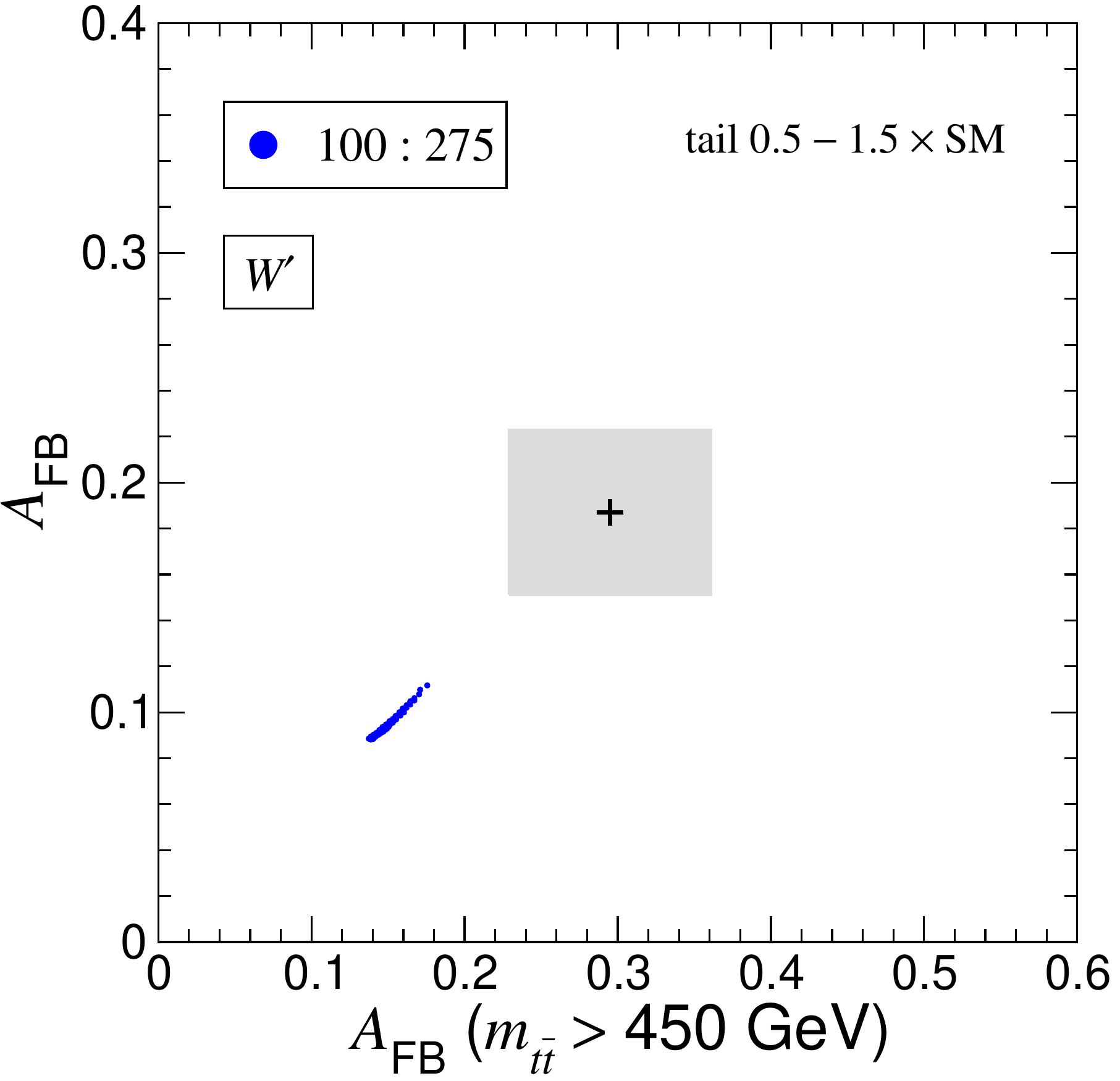} &
\includegraphics[height=4.8cm]{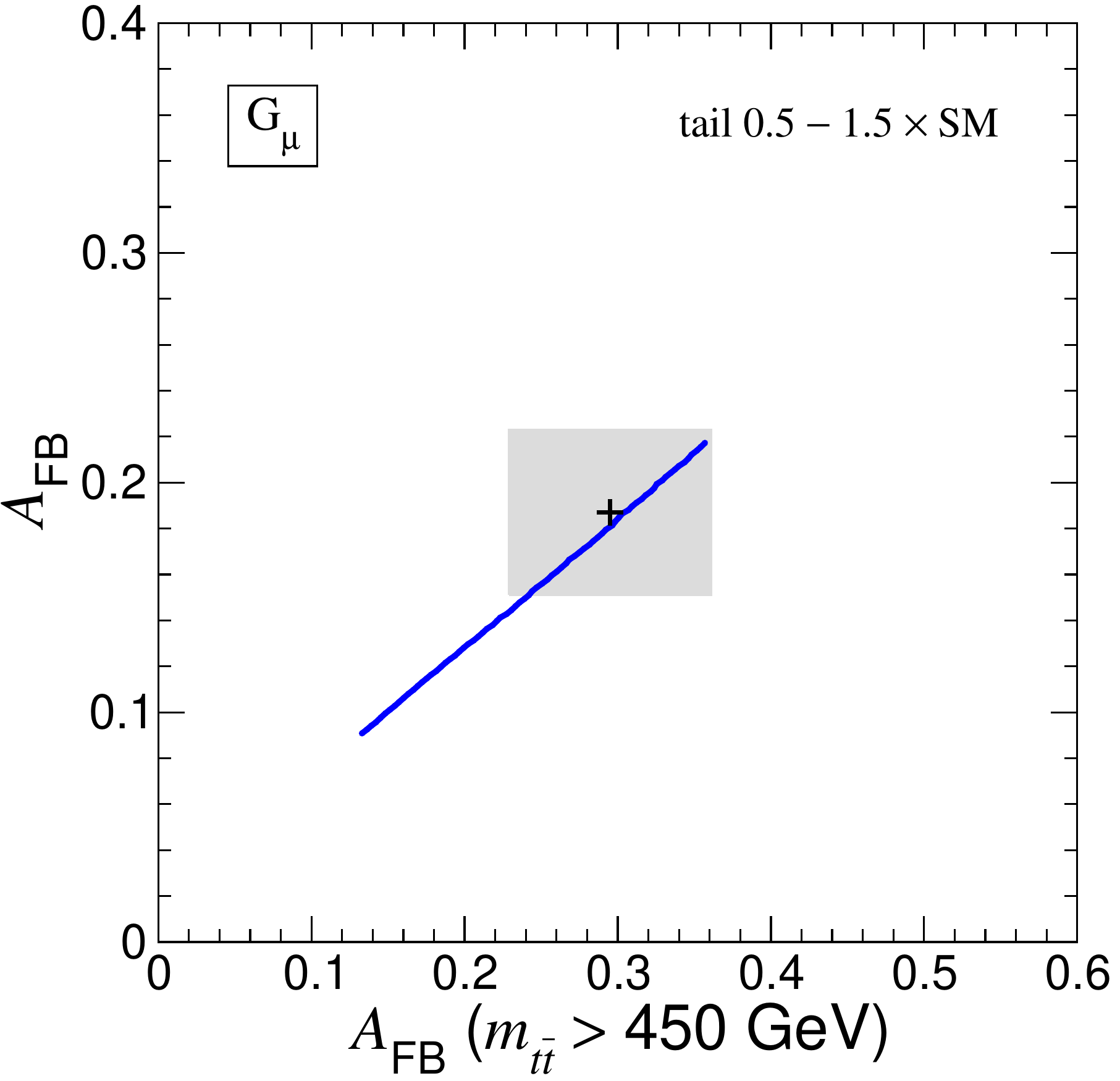} \\
\includegraphics[height=4.8cm]{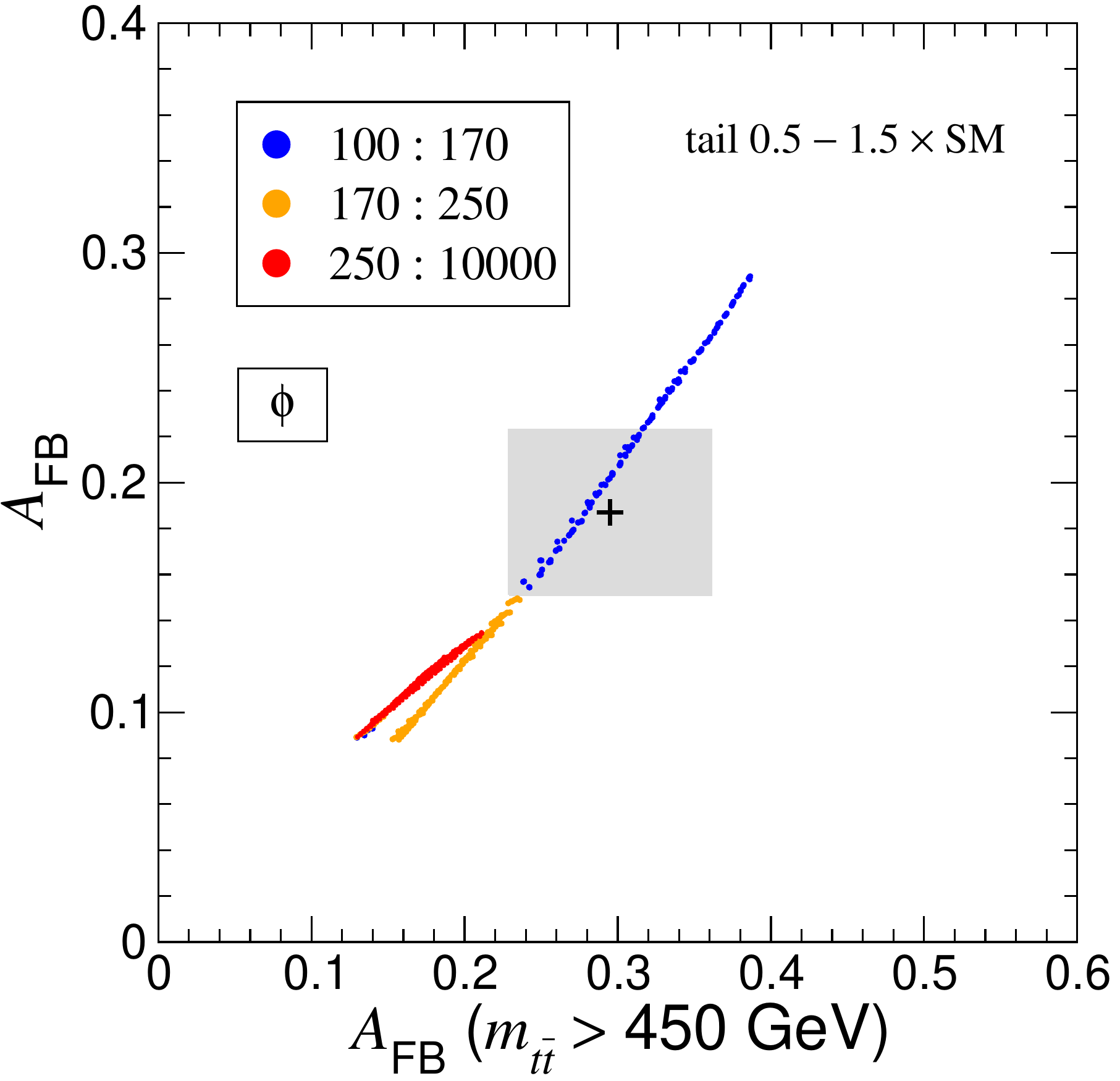} & \includegraphics[height=4.8cm]{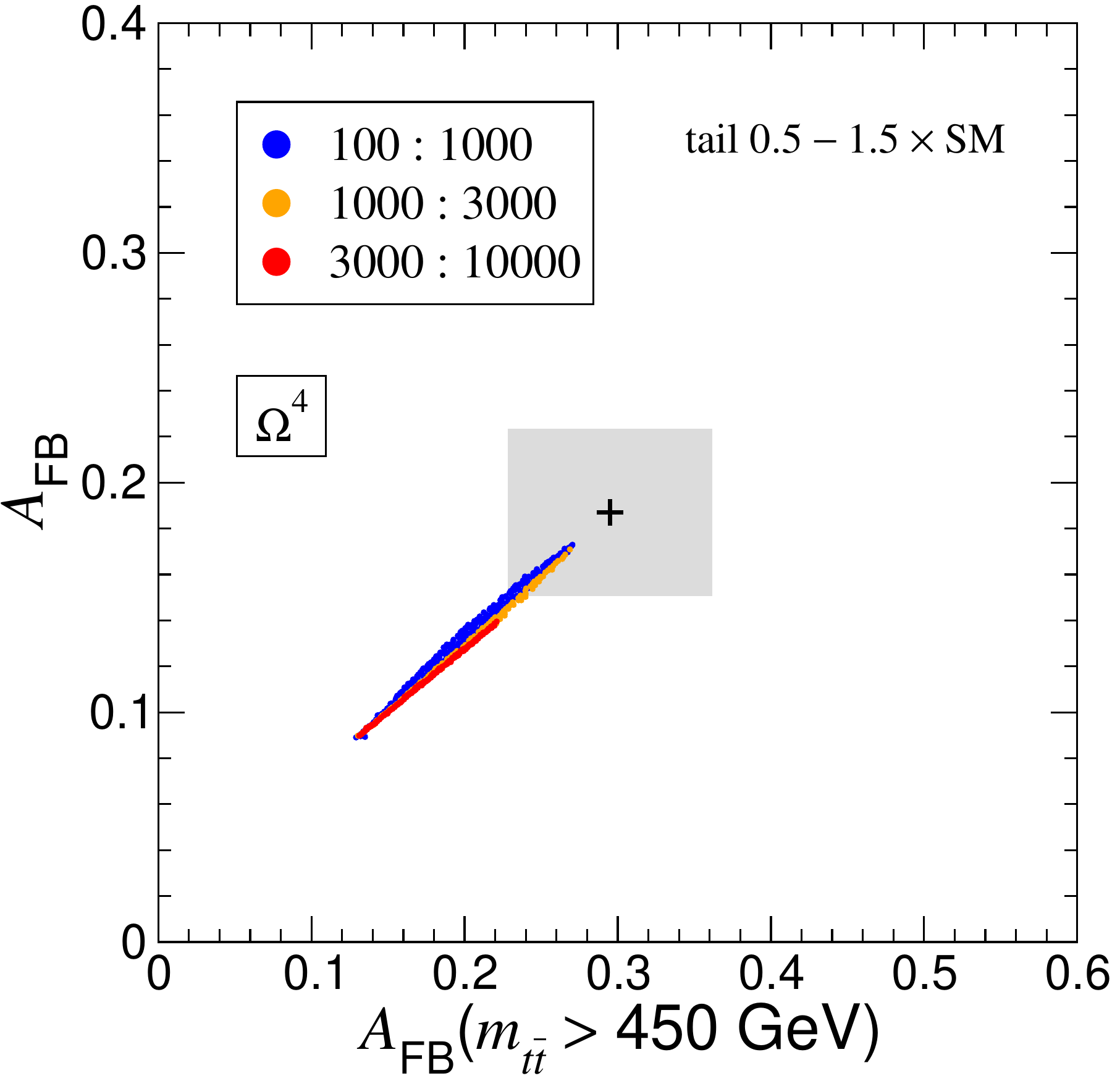} &
\includegraphics[height=4.8cm]{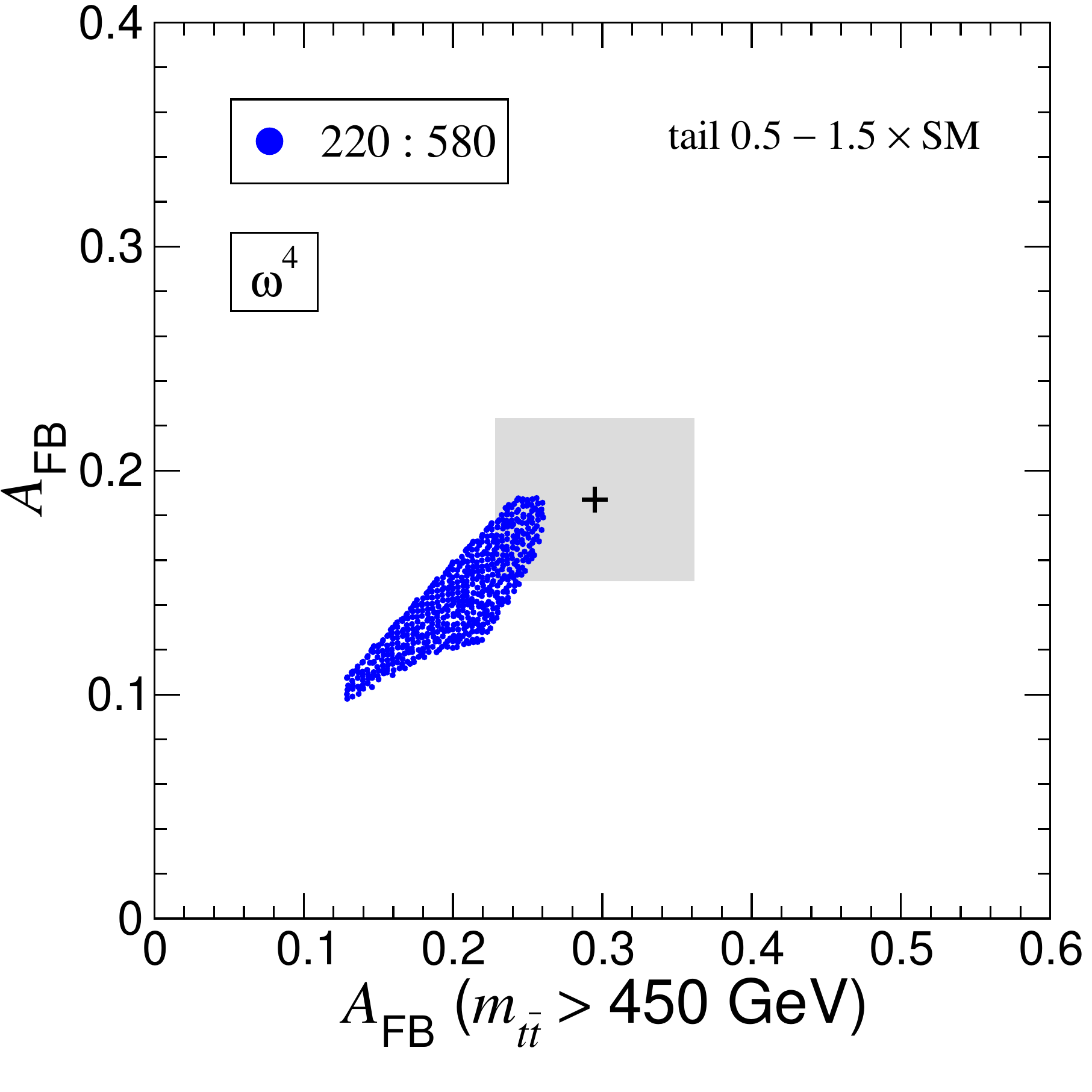}
\end{tabular}
\end{center}
\caption{\label{fig:AvsAHi}Inclusive versus high-mass asymmetries at the Tevatron, for several new physics models, imposing a tight constraint on the LHC high-mass tail (see the text).The numbers in the legends indicate the mass range for the new particle, in GeV. The crosses correspond to the experimental measurements in in Fig.~\ref{fig:tevX}, with the shaded boxes corresponding to the $1\sigma$ uncertainty.}
\end{figure}

One of the first and most universal predictions for models explaining the Tevatron excess is an enhancement of the $t \bar t$ differential distribution at high $m_{t \bar t}$~\cite{AguilarSaavedra:2011vw,Delaunay:2011gv}. The LHC experiments have not found any sign of tail enhancement~\cite{Aad:2012hg,:2012qka}. The most stringent limits result from the ATLAS analysis, which measures a cross section slightly below the SM prediction, $d\log \sigma/dm_{t \bar t} \simeq 7 \pm 2~\mathrm{PeV}^{-1}$  for $m_{t \bar t} \geq 950$ GeV, compared to a SM prediction of $9~\mathrm{PeV}^{-1}$. This leads to an upper limit $\sigma/\sigma_\mathrm{SM} \leq 1.3$ in this mass bin, with a 99\% confidence level. With all caveats that apply to a naive interpretation of an experimental result, we observe that this limit is similar to the one projected in Ref.~\cite{AguilarSaavedra:2011ug}, $\sigma/\sigma_\mathrm{SM} \leq 1.5$ for $m_{t \bar t} \geq 1$ TeV. Imposing the latter constraint as well as a minimum of $\sigma/\sigma_\mathrm{SM} \geq 0.5$, the allowed areas for the Tevatron asymmetries in Fig.~\ref{fig:AvsAH} shrink to the corresponding ones in Fig.~\ref{fig:AvsAHi}.
The $Z'$ model, already discarded because it overpredicts the Tevatron excess, does not give a positive contribution to the Tevetron asymmetry after imposing the LHC tail constraint. The $W'$ model, which predicted too large values for $A_C$, cannot reproduce the Tevatron asymmetries either. The rest of models are, in principle, compatible with $t \bar t$ cross section measurements at the LHC and the Tevatron.
Among them, the ``least disturbing'' one for the differential distribution is an $s$-channel colour octet, as long as it is heavy (or light) enough. However, if the resonance is at kinematical reach the differential cross section enhancement is much larger, and the new particle will appear as a peak (or bump, if it is very wide) in the $t \bar t$ invariant mass spectrum. Current LHC searches put heavy gluon models into trouble as the mass scales probed go higher, because the couplings required get too large and nonperturbative. However, these constraints can be avoided by ``light'' gluons with a mass of few hundreds of GeV~\cite{Barcelo:2011vk,Tavares:2011zg,AguilarSaavedra:2011ci,Krnjaic:2011ub}.

\section{A new friend: the top polarisation} 

Another common signal of new physics contributions to $t \bar t$ production, which in general may couple differently to $t_L$ and $t_R$, is a change in the polarisation of the produced top (anti)quarks. The double angular distribution for the production of a $t \bar t$ pair is
\begin{equation}
\frac{1}{\sigma} \frac{d\sigma}{d\cos \theta_t \, d\cos \theta_{\bar t}} = \frac{1}{4} \left[ 1+B_t \cos \theta_t + B_{\bar t} \cos \theta_{\bar t} + C \cos \theta_t \cos \theta_{\bar t} \right] \,.
\end{equation}
with $\theta_t$, $\theta_{\bar t}$ the angles between the top (antitop) momenta in the zero momentum frame (ZMF) with respect to some chosen spin axes~\cite{Bernreuther:2004jv,Bernreuther:2010ny}. The constants $B_t$, $B_{\bar t}$ correspond to the polarisation of the top and antitop, respectively.\footnote{Provided CP is conserved, $B_{\bar t}=-B_t$ if the same axis is chosen to measure the top and antitop spins. Choosing the helicity basis for each quark, $B_t = B_{\bar t}$.} In the SM they vanish at the tree level ---that is, top (anti-)quarks are produced unpolarised--- due to the vector structure of the QCD coupling, and they are small at higher orders. The $C$ constant measures the spin correlation between the top and antitop, and is nonzero for a suitable choice of spin axes.

\begin{figure}[htb]
\begin{center}
\begin{tabular}{ccc}
\includegraphics[height=6cm]{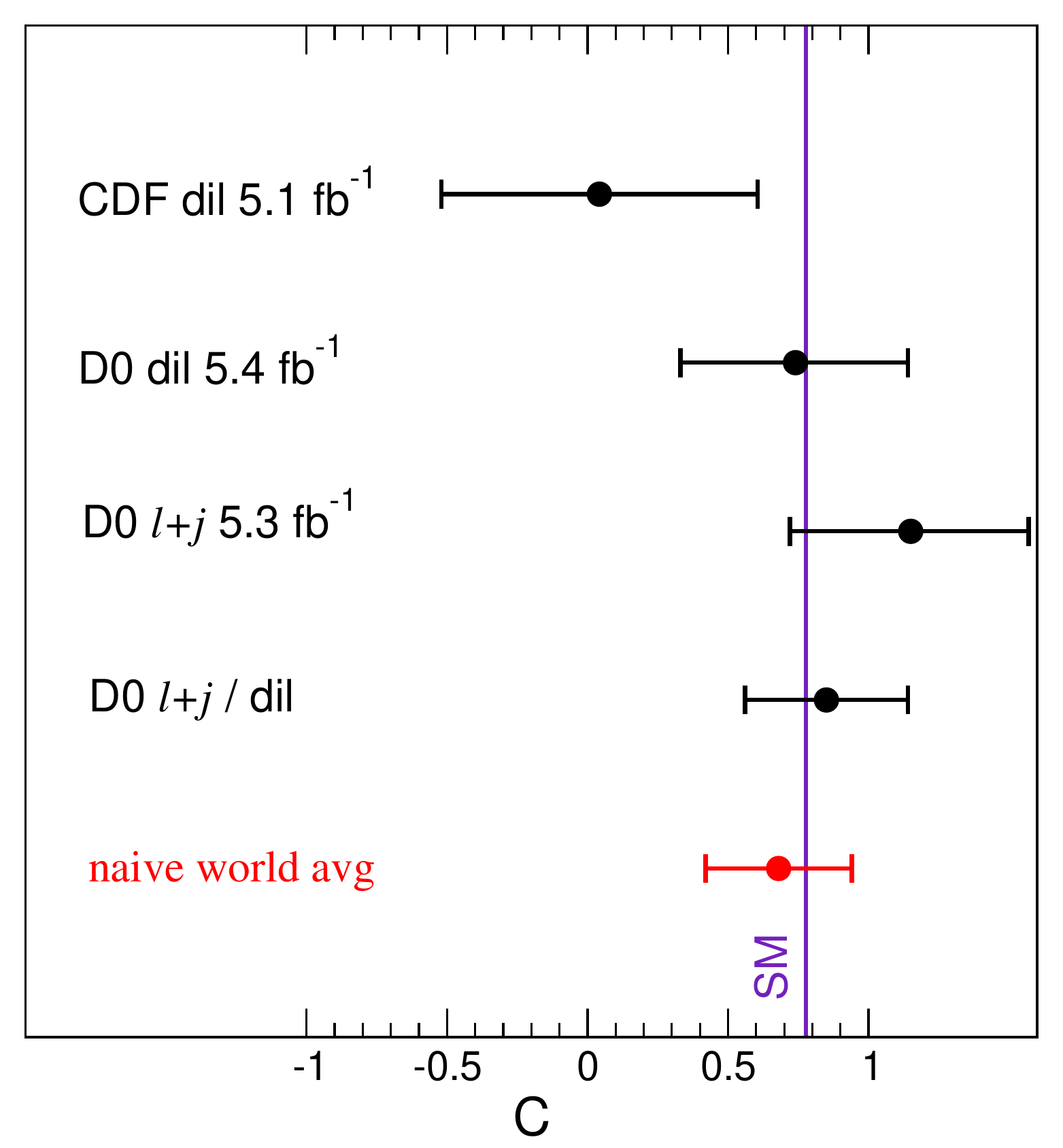} & \quad
\raisebox{5mm}{\includegraphics[height=5.3cm]{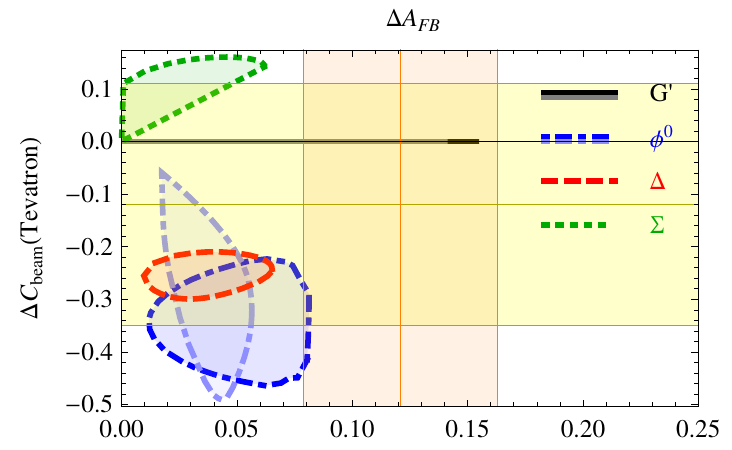}} \\[2mm]
\includegraphics[height=6cm]{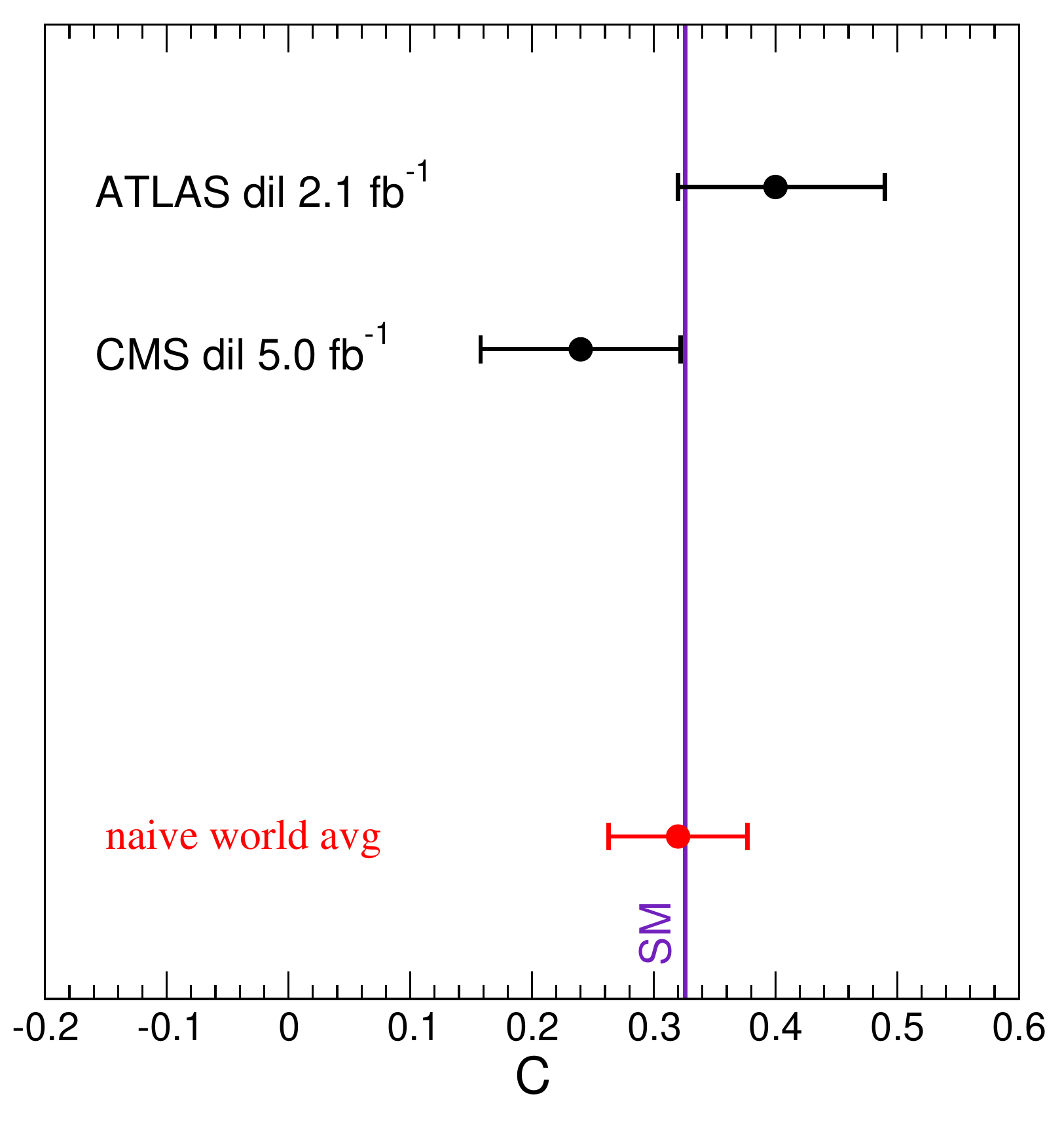} & \quad
\raisebox{5mm}{\includegraphics[height=5.3cm]{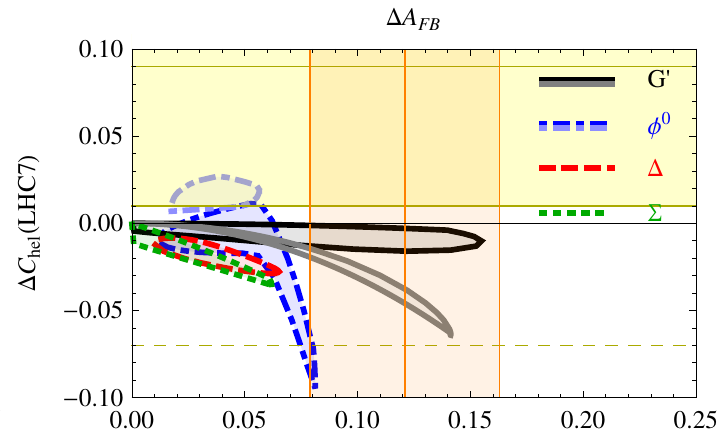}} \\[2mm]
\includegraphics[height=6cm]{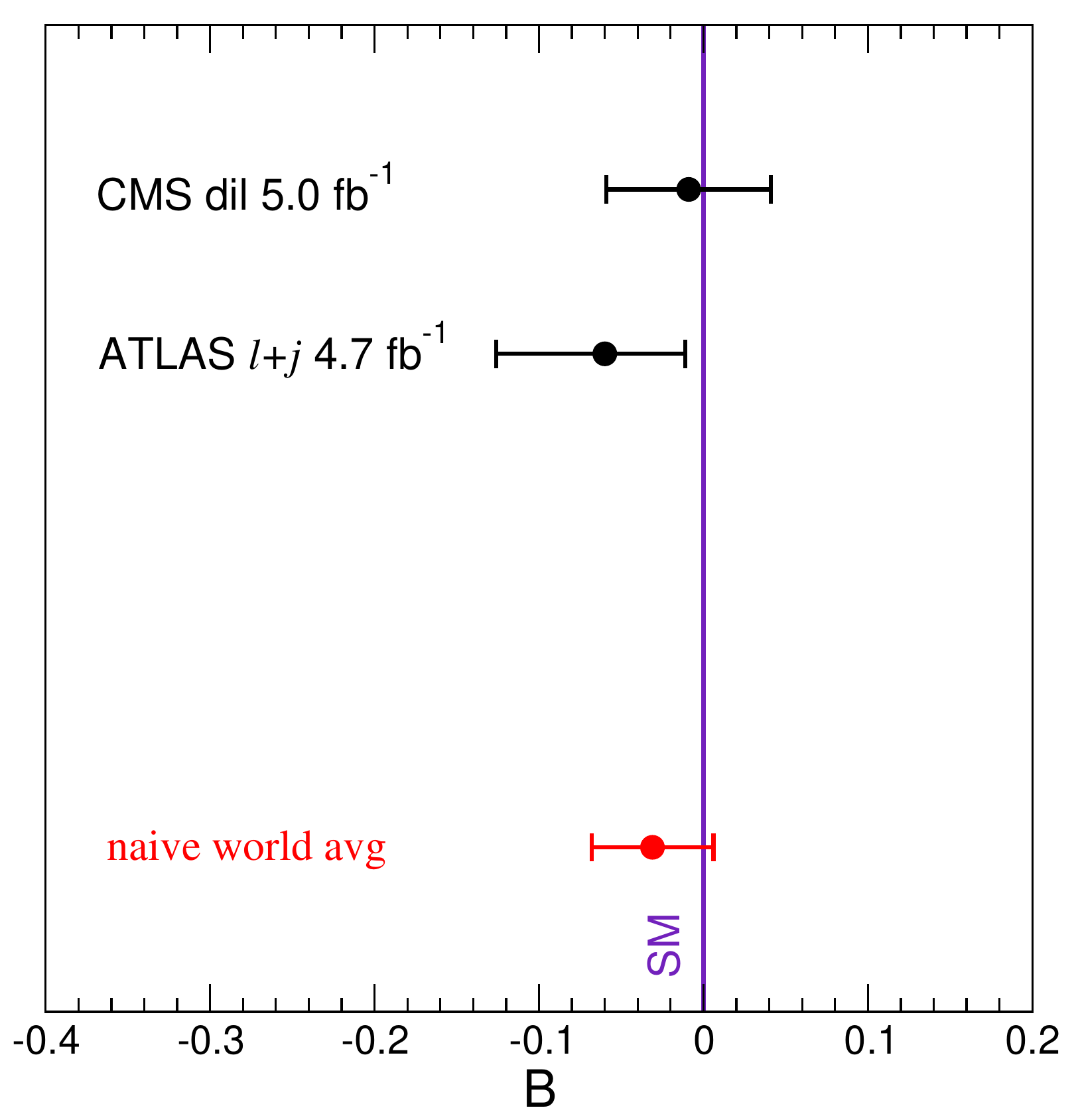} & \quad
\raisebox{5mm}{\includegraphics[height=5.3cm]{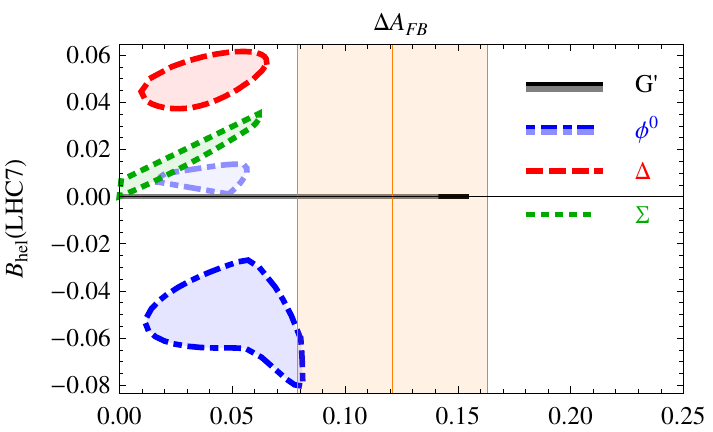}}
\end{tabular}
\end{center}
\caption{\label{fig:pol}Left: experimental measurements of spin observables. Right (taken from Ref.~\cite{Fajfer:2012si}): theoretical predictions within several new physics models. The horizontal axis represents the increase in $\afb$ from new physics sources and the vertical axis the increase in the corresponding ($B$ or $C$) coefficient, with respect to the SM value. The legends $\Delta$, $\Sigma$ correspond to $\omega^4$, $\Omega^4$ in our notation.}
\end{figure}

The spin correlation coefficient $C$ has been measured at the Tevatron, in the ``beamline'' basis~\cite{CDF-C,Abazov:2011ka,Abazov:2011gi} (Fig.~\ref{fig:pol}, upper left panel), giving a naive average of $C=0.68 \pm 0.26$, in good agreement with the SM prediction $C_\mathrm{SM}=0.79$~\cite{Bernreuther:2010ny}. Defining $\Delta C = C-C_\mathrm{SM}$, the experimental measurement $\Delta C = -0.11 \pm 0.26$ already sets some constraints on the possible contributions to $\afb$ from new physics~\cite{Fajfer:2012si} (Fig.~\ref{fig:pol}, upper right panel).

At the LHC, the spin correlation has been determined in the helicity basis~\cite{ATLAS:2012ao,CMS:2012cxa}, with an average measurement $C=0.32 \pm 0.06$ (Fig.~\ref{fig:pol}, middle left panel). Defining again $\Delta C = C-C_\mathrm{SM}$, with $C_\text{SM} = 0.31$~\cite{Bernreuther:2010ny}, this average corresponds to $\Delta C = 0.01 \pm 0.06$. The measurement of $\Delta C$ at the LHC has little effect on the allowed parameter space for models reproducing the Tevatron $\afb$ (Fig.~\ref{fig:pol}, middle right panel). More restrictive is the measurement of the top polarisation, {\it i.e.} the $B_t$ parameter~\cite{CMS:2012owa,ATLAS:2012epa} (Fig.~\ref{fig:pol}, lower left panel). The average of the ATLAS and CMS measurements, $B=-0.03 \pm 0.04$, disfavours at the $\sim 2\sigma$ level the explanation of the asymmetry by colour sextet and triplet scalars, which couple to $t_R$ and predict a positive top quark polarisation (Fig.~\ref{fig:pol}, lower right panel).

\begin{figure}[htb]
\begin{center}
\includegraphics[height=6cm]{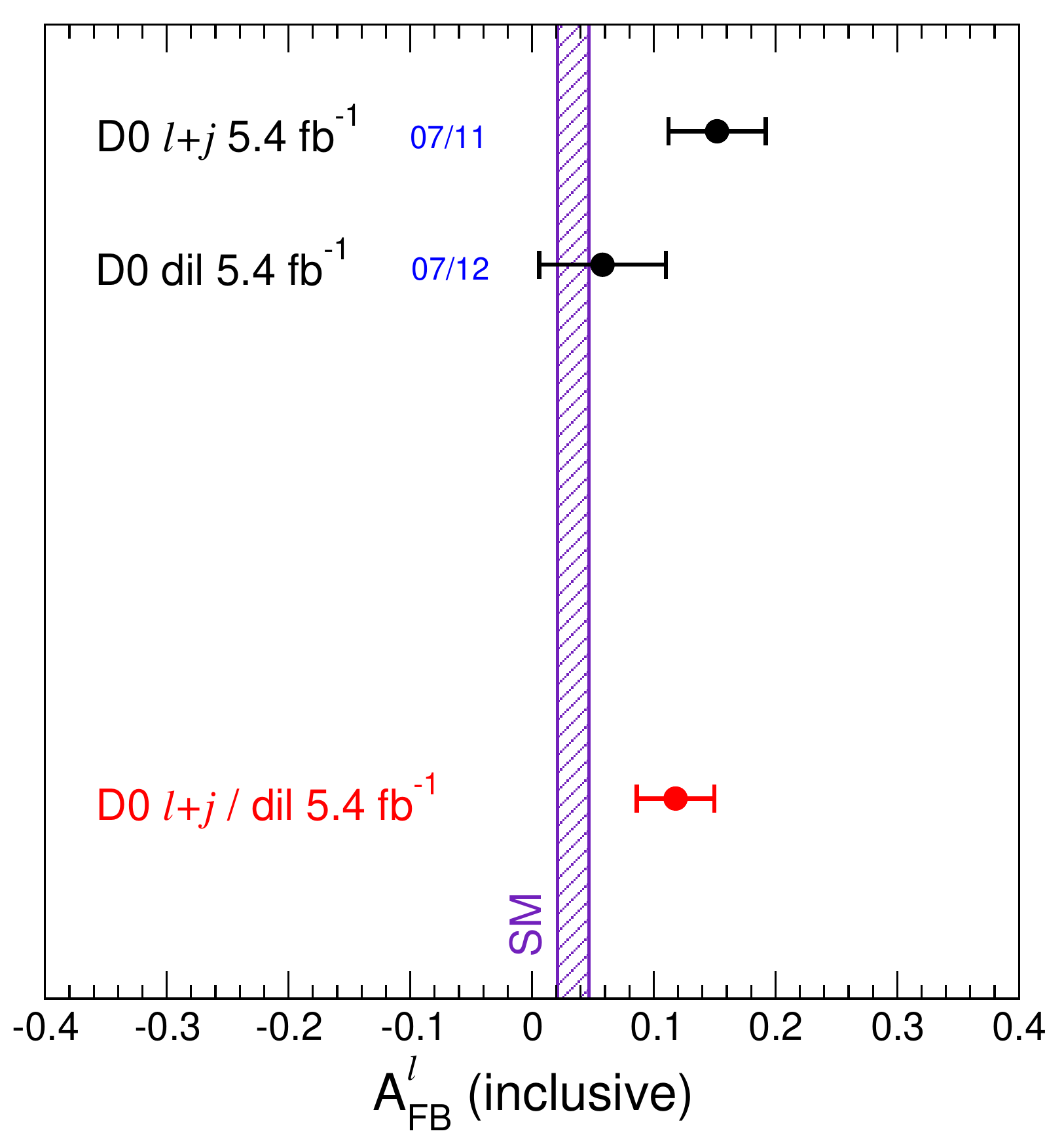}
\end{center}
\caption{\label{fig:alFB}Measurements of the lepton asymmetry $A_{FB}^\ell$ at the Tevatron.}
\end{figure}

For completeness, it is worth mentioning that lepton-based FB asymmetries have also been measured at the Tevatron~\cite{Abazov:2011rq,:2012bfa},
\begin{eqnarray}
A_{FB}^\ell & = &  \frac{N(Q \cdot \eta>0) - N(Q \cdot \eta<0)}{N(Q \cdot \eta>0) + N(Q \cdot \eta<0)} \,, \nonumber \\
A^{\ell \ell} & = &  \frac{N(\Delta \eta>0) - N(\Delta \eta<0)}{N(\Delta \eta>0) + N(\Delta \eta<0)} \,,
\end{eqnarray}
with $\eta$ the rapidity of the charged leptons and $Q$ their charge.
These asymmetries include information from the $t \bar t$ FB asymmetry and the top polarisation~\cite{Berger:2012nw,Falkowski:2012cu} ---the latter is found in agreement with the SM--- and provide a complementary experimental handle to probe new contributions to $t \bar t$ production. The measurements of $A_{FB}^\ell$ (see Fig.~\ref{fig:alFB}) are more precise and exhibit a positive excess with respect to SM predictions, whereas the only measurement of the dilepton asymmetry, $A^{\ell \ell} = 0.053 \pm 0.084$, is compatible with the SM prediction $A_\mathrm{SM}^{\ell \ell} = 0.062$~\cite{Bernreuther:2012sx}.

\section{Discussion}

More than two years after the measurement~\cite{Aaltonen:2011kc} that trigged the wide interest in the FB asymmetry at the Tevatron, there have been various new measurements, updated SM predictions and plenty of proposals for new physics explanations. But the question still remains whether the Tevatron anomaly corresponds to new physics, an unknown higher-order SM correction, or some kind of systematic effect.

\begin{figure}[htb]
\begin{minipage}{7.5cm}
\begin{center}
\includegraphics[height=5cm]{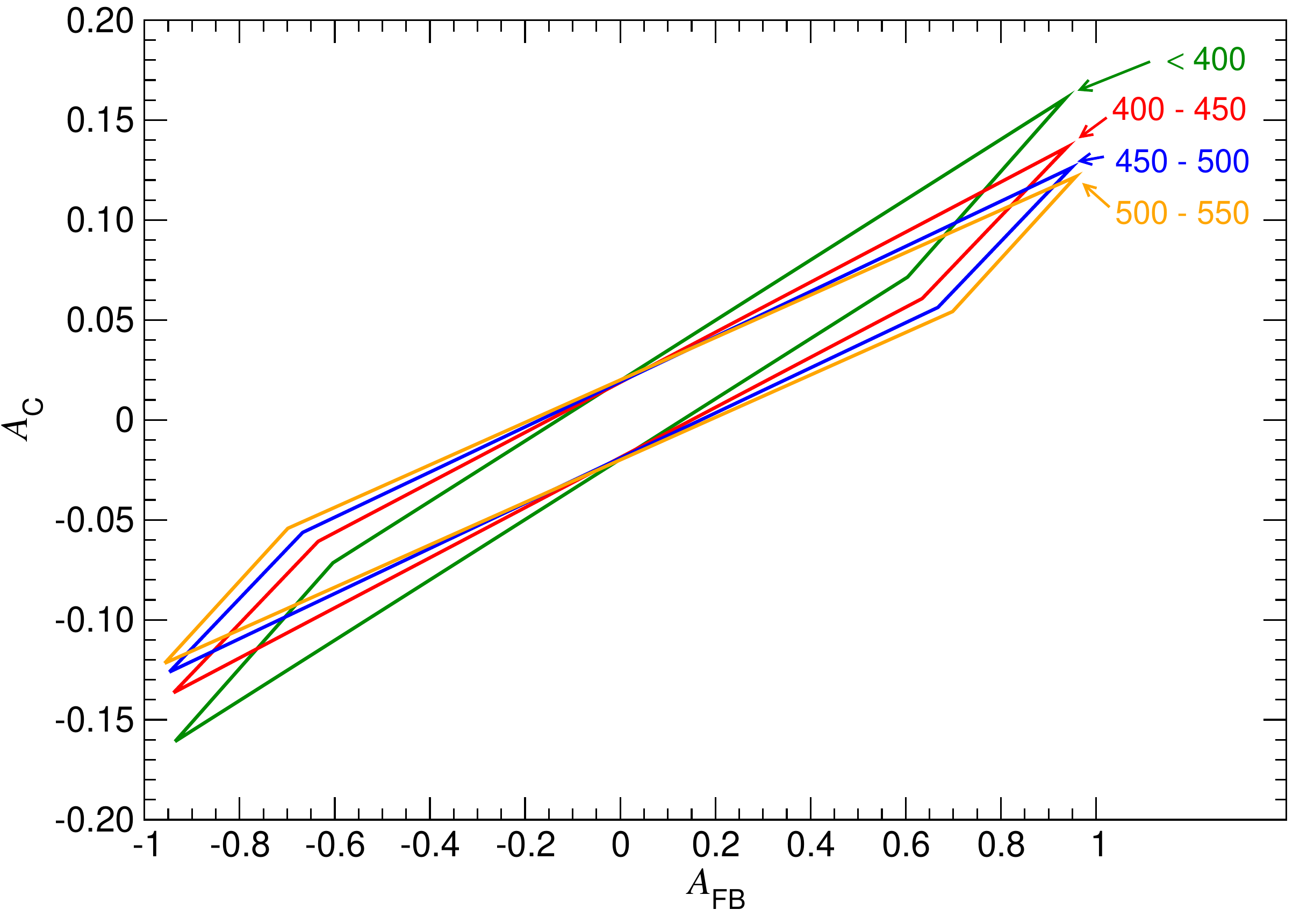}
\caption{\label{fig:Apred1}Model-independent prediction for the correlation between $\ac$ and $\afb$ in several $m_{t \bar t}$ bins, based on the collider-independent asymmetries.}
\end{center}
\end{minipage}\hspace{2pc}%
\begin{minipage}{7.5cm}
\begin{center}
\includegraphics[height=5.5cm]{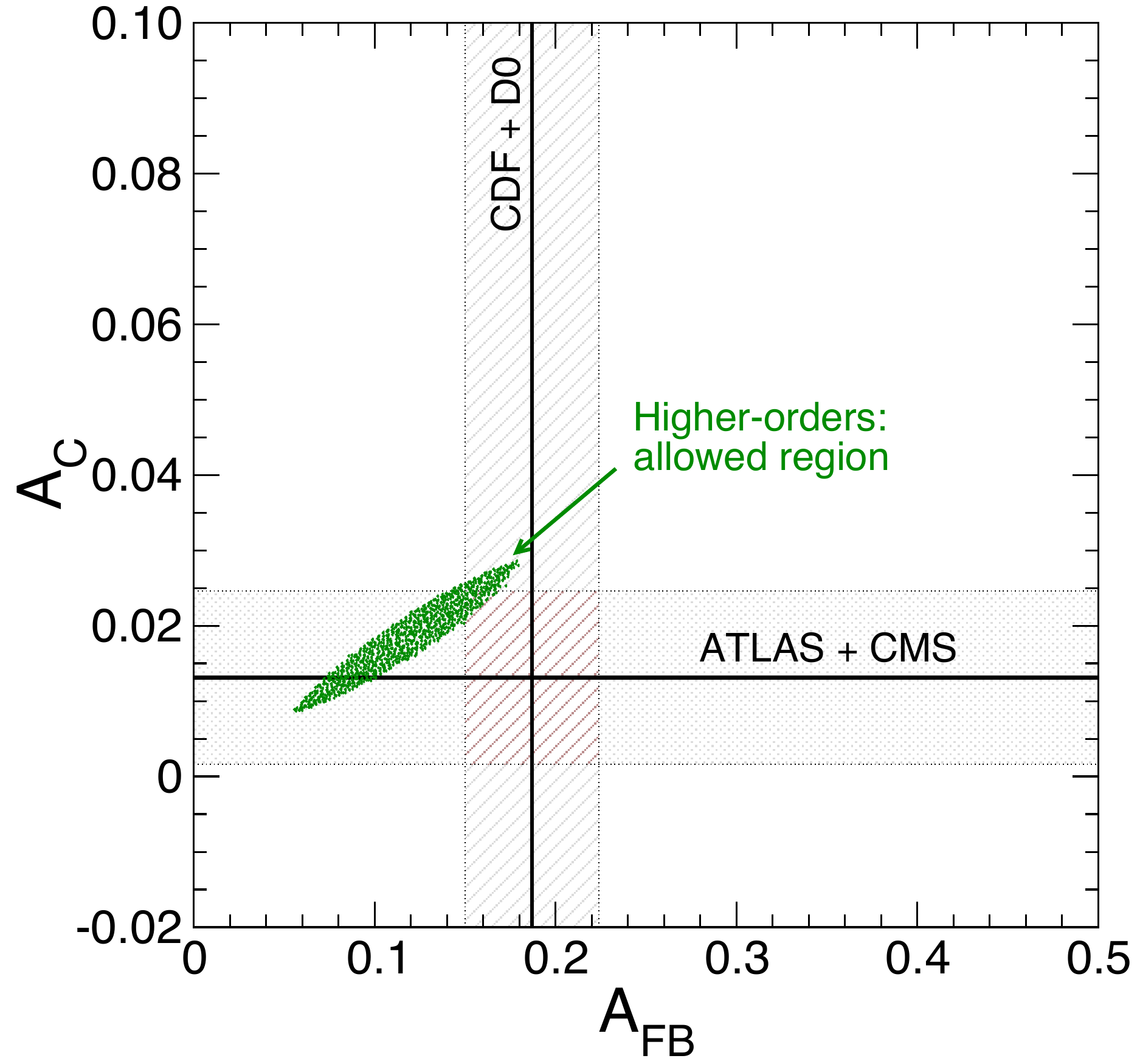}
\caption{\label{fig:Apred2}Estimated correlation between $\ac$ and $\afb$ from higher-order QCD effects, based on the collider-independent asymmetries.}
\end{center}
\end{minipage} 
\end{figure}

The first possibility, of new physics in $t \bar t$ production, is certainly exciting, but it is not clear which form this new physics could have. Among the six simple models that we considered as candidates to explain the excess in the Tevatron asymmetry, only two of them (a colour octet and a scalar doublet) have survived after imposing just constraints from other observables in $t \bar t$ production at the Tevatron and LHC. But, in addition, there are some other measurements that put the minimal implementation of these models in trouble.
\begin{itemize}
\item Light gluons mediate dijet pair~\cite{Gross:2012bz} and four-top production~\cite{AguilarSaavedra:2011ck}, neither observed. The former constitute a serious constraint for lighter masses, which can be softened by assuming a large gluon width ---due for example to decays into some new particle. The latter may become relevant for a wide mass range with increased luminosity. In addition, there are constraints from low-energy $B$ physics~\cite{Bai:2011ed,Ipek:2013zi} and electroweak precision data~\cite{Gresham:2012kv} that are very model-dependent. 
\item For scalar doublets there are also stringent constraints from $B$ physics~\cite{Blum:2011fa}, atomic parity violation~\cite{Gresham:2012wc} and possibly from associated production with a top quark.
\end{itemize}
These difficulties, in any case, may just reflect our current inability to propose a compelling model that explains the asymmetry measurements at the Tevatron and the LHC, fulfilling the constraints from other $t \bar t$ observables and collider data without ad-hoc assumptions and model fine-tuning. In this regard, we have stressed that a Tevatron $\afb$ excess and a small $\ac$ at the LHC are compatible in general. By using Eqs.~(\ref{ec:AuAd}) and considering $A_u$, $A_d$ as free independent parameters ranging between $-1$ and $1$, one can obtain predictions for the correlated asymmetries $\afb$, $\ac$ within each $m_{t \bar t}$ bin. This is shown in Fig.~\ref{fig:Apred1}. A positive excess at the Tevatron is compatible with a zero, or even negative, asymmetry at the LHC, provided there is some cancellation between the contributions from $A_u$ and $A_d$.

Explanations of the Tevatron excess by higher-order QCD effects are unlikely to fit data well, if the trend of current measurements persists. A simple reckoning indicates that, if there were large missing QCD corrections that shifted $\afb$ to its Tevatron average, they would also shift $\ac$ away from the LHC average. In the absence of a proper next-to-next-to-leading order calculation, one can give a back-of-the-envelope estimate for this correlation by using again Eqs.~(\ref{ec:AuAd}) and varying $A_u$, $A_d$ around the NLO SM value.\footnote{We thank W. Bernreuther and Z.-G. Si for providing us with these data.} The resulting prediction for $\ac$ versus $\afb$ in Fig.~\ref{fig:Apred2} shows that one cannot simultaneously fit the central values of Tevatron and LHC asymmetries with this kind of corrections to the SM. A similar type of argument is expected to hold for other QCD-based explanations of the anomalies~\cite{Skands:2012mm,Brodsky:2012sz}.

The third option, of unknown systematic errors in Tevatron or LHC experiments, is hard to understand since the two experiments at each collider provide similar results. Still, unknown systematics are unknown by definition, and little can be said in this direction at the moment.

In summary, the $\afb$ puzzle is far from being solved. There are still good hopes that there is some type of new physics in $t \bar t$ production, which might (or might not) be visible by precision measurements of the LHC charge asymmetry, the $t \bar t$ differential distributions and the top polarisation. Fortunately, the upcoming LHC and D0 measurements will provide important information to help approach a solution.\footnote{Definitely, {\it ``When you have eliminated the impossible, whatever remains, however improbable, must be the truth.''}~\cite{SH}.}

\section*{Acknowledgements}
This work has been supported by MICINN by projects FPA2006-05294 and FPA2010-17915, Junta de Andaluc\'{\i}a (FQM 101, FQM 03048 and FQM 6552) and Funda\c c\~ao
para a Ci\^encia e Tecnologia~(FCT) project CERN/FP/123619/2011.


\section*{References}
\begin{thebibliography}{99}
\bibitem{Aaltonen:2012it}
  Aaltonen T {\it et al.}  (CDF Collaboration) 2012
  {\it Preprint} arXiv:1211.1003 [hep-ex].

\bibitem{Abazov:2011rq}
  Abazov V M {\it et al.} (D0 Collaboration) 2011
  {\it Phys.\ Rev.}\ D {\bf 84} 112005.

\bibitem{Campbell:2012uf}
  Campbell J M and Ellis R K 2012
  {\it Preprint} arXiv:1204.1513 [hep-ph].

\bibitem{Ahrens:2011uf}
  Ahrens V, Ferroglia A, Neubert M, Pecjak B D and Yang L L 2011
  {\it Phys.\ Rev.}\ D {\bf 84} 074004.

\bibitem{Hollik:2011ps}
  Hollik W and Pagani D 2011
  {\it Phys.\ Rev.}\ D {\bf 84} 093003.

\bibitem{Kuhn:2011ri}
  Kuhn J H and Rodrigo G 2012
  {\it JHEP} {\bf 1201} 063.

\bibitem{Bernreuther:2012sx}
  Bernreuther W and Si Z G 2012
  {\it Phys.\ Rev.}\ D {\bf 86} 034026.

\bibitem{Aaltonen:2011kc}
  Aaltonen T {\it et al.}  (CDF Collaboration) 2011
  {\it Phys.\ Rev.}\ D {\bf 83} 112003.

\bibitem{AguilarSaavedra:2011vw}
  Aguilar-Saavedra J A and P\'erez-Victoria M 2011
  {\it JHEP} {\bf 1105} 034.

\bibitem{Djouadi:2009nb} 
  Djouadi A, Moreau G, Richard G and Singh R K 2010
  {\it Phys.\ Rev.}\ D {\bf 82}, 071702.

\bibitem{Jung:2009jz} 
  Jung S, Murayama H, Pierce A and Wells J D 2010
  {\it Phys.\ Rev.}\ D {\bf 81}, 015004.

\bibitem{Cheung:2009ch}
  Cheung K, Keung W Y and Yuan T C 2009
  {\it Phys.\ Lett.}\ B {\bf 682} 287.

\bibitem{Nelson:2011us} 
  Nelson A E, Okui T and Roy T S 2011
  {\it Phys.\ Rev.}\ D {\bf 84}, 094007.

\bibitem{Shu:2009xf} 
  Shu J, Tait T M P and Wang K 2010
  {\it Phys.\ Rev.}\ D {\bf 81}, 034012.

\bibitem{Gabrielli:2011jf}
  Gabrielli E and Raidal M 2011
  {\it Phys.\ Rev.}\ D {\bf 84} 054017.

\bibitem{Aliev:2010zk}
  Aliev M, Lacker H, Langenfeld U, Moch S, Uwer P and Wiedermann M 2011
  {\it Comput.\ Phys.\ Commun.}\  {\bf 182} 1034.

\bibitem{ttxsexp}
Aaltonen T {\it et al.} (CDF Collaboration) 2012 {\it CDF note 10926}; Abazov V M {\it et al.} (D0 Collaboration) 2012 {\it D0 Note 6363-CONF}.


\bibitem{Grinstein:2011yv}
  Grinstein B, Kagan A L, Trott M and Zupan J 2011
  {\it Phys.\ Rev.\ Lett.}\  {\bf 107} 012002.


\bibitem{AguilarSaavedra:2011ug}
  Aguilar-Saavedra J A and P\'erez-Victoria M 2011
  {\it JHEP} {\bf 1109} 097.

\bibitem{AguilarSaavedra:2010zi}
  Aguilar-Saavedra J A 2011
  {\it Nucl.\ Phys.}\ B {\bf 843} 638

\bibitem{Degrande:2010kt}
  Degrande C, Gerard J M, Grojean C, Maltoni F and Servant G 2011
  {\it JHEP} {\bf 1103} 125.

\bibitem{AguilarSaavedra:2011zy}
  Aguilar-Saavedra J A and P\'erez-Victoria M 2011 2011
  {\it Phys.\ Lett.}\ B {\bf 701} 93.

\bibitem{AguilarSaavedra:2011ci}
  Aguilar-Saavedra J A and P\'erez-Victoria M 2011
  {\it Phys.\ Lett.}\ B {\bf 705} 228.

\bibitem{CDF-theta}
  Aaltonen T {\it et al.} (CDF Collaboration), {\it CDF note 10947}.

\bibitem{Gresham:2011pa}
  Gresham M I, Kim I W and Zurek K M 2011
  {\it Phys.\ Rev.}\ D {\bf 83} 114027.

\bibitem{Diener:2009ee}
  Diener R, Godfrey A and Martin T A W 2009
  {\it Phys.\ Rev.}\ D {\bf 80} 075014.

\bibitem{Frixione:2002ik}
  Frixione S and Webber B R 2002
  {\it JHEP} {\bf 0206} 029.

\bibitem{AguilarSaavedra:2012va}
  Aguilar-Saavedra J A and Juste A 2012
  {\it Phys.\ Rev.\ Lett.}\  {\bf 109} 211804.

\bibitem{AguilarSaavedra:2011hz}
  Aguilar-Saavedra J A and P\'erez-Victoria M 2011
  {\it Phys.\ Rev.}\ D {\bf 84} 115013.

\bibitem{Ko:2012ud}
  Ko P, Omura Y and Yu C 2012
  {\it Preprint} arXiv:1205.0407 [hep-ph].

\bibitem{Drobnak:2012cz}
  Drobnak J, Kamenik J F and Zupan J 2012
  {\it Phys.\ Rev.}\ D {\bf 86} 054022.

\bibitem{Drobnak:2012rb}
  Drobnak J, Kagan A L, Kamenik J F, Perez G and Zupan J 2012
  {\it Phys.\ Rev.}\ D {\bf 86} 094040.

\bibitem{Alvarez:2012ca}
  Alvarez E and Leskow E C 2012
  {\it Phys.\ Rev.}\ D {\bf 86} 114034.

\bibitem{AguilarSaavedra:2012rx}
  Aguilar-Saavedra J A, Bernreuther W and Si Z G 2012
  {\it Phys.\ Rev.}\ D {\bf 86} 115020.

\bibitem{AguilarSaavedra:2011cp}
  Aguilar-Saavedra J A, Juste A and Rubbo F 2012
  {\it Phys.\ Lett.}\ B {\bf 707} 92.

\bibitem{Delaunay:2011gv}
  Delaunay C, Gedalia O, Hochberg Y, Perez G and Soreq Y 2011
  {\it JHEP} {\bf 1108} 031.

\bibitem{Aad:2012hg}
  The ATLAS Collaboration 2012
  {\it Preprint} arXiv:1207.5644 [hep-ex].

\bibitem{:2012qka}
  The CMS Collaboration 2012
  {\it Preprint} arXiv:1211.2220 [hep-ex].

\bibitem{Barcelo:2011vk}
  Barcelo R, Carmona A, Masip M and Santiago J 2012
  {\it Phys.\ Lett.}\ B {\bf 707} 88.

\bibitem{Tavares:2011zg}
  Marques Tavares G and Schmaltz M 2011
  {\it Phys.\ Rev.}\ D {\bf 84} 054008.

\bibitem{Krnjaic:2011ub}
  Krnjaic G Z 2012
  {\it Phys.\ Rev.}\ D {\bf 85} 014030.

\bibitem{Bernreuther:2004jv}
  Bernreuther W, Brandenburg A, Si Z G and Uwer P 2004
  {\it Nucl.\ Phys.}\ B {\bf 690} 81.
  
\bibitem{Bernreuther:2010ny}
  Bernreuther W and Si Z G 2010
  {\it Nucl.\ Phys.}\ B {\bf 837} 90.

\bibitem{CDF-C}
  Aaltonen T {\it et al.} (CDF Collaboration), {\it CDF note 10719}.

\bibitem{Abazov:2011ka}
  Abazov V M {\it et al.}  (D0 Collaboration) 2011
  {\it Phys.\ Rev.\ Lett.}\  {\bf 107} 032001,

\bibitem{Abazov:2011gi}
  Abazov V M {\it et al.} (D0 Collaboration) 2012
  {\it Phys.\ Rev.\ Lett.}\  {\bf 108} 032004.

\bibitem{Fajfer:2012si}
  Fajfer S, Kamenik J F and Melic B 2012
  {\it JHEP} {\bf 1208} (2012) 114.

\bibitem{ATLAS:2012ao}
  The ATLAS Collaboration 2012
  {\it Phys.\ Rev.\ Lett.}\  {\bf 108} 212001.

\bibitem{CMS:2012cxa}
  The CMS Collaboration 2012
  {\it CMS-PAS-TOP-12-004}.

\bibitem{CMS:2012owa}
  The CMS Collaboration 2012
  {\it CMS-PAS-TOP-12-016}.

\bibitem{ATLAS:2012epa}
  The ATLAS Collaboration 2012
  {\it ATLAS-CONF-2012-133}.

\bibitem{:2012bfa}
  Abazov V M {\it et al.}  (D0 Collaboration) 2012
  {\it Preprint} arXiv:1207.0364 [hep-ex].

\bibitem{Berger:2012nw}
  Berger E L, Cao Q H, Chen C R, Yu J H and Zhang H 2012
  {\it Phys.\ Rev.\ Lett.}\  {\bf 108} 072002.

\bibitem{Falkowski:2012cu}
  Falkowski A, Mangano M L, Martin A, Perez G and Winter J 2012
  {\it Preprint} arXiv:1212.4003 [hep-ph].

\bibitem{Gross:2012bz}
  Gross C, Marques Tavares G, Schmaltz M and Spethmann C 2013
  {\it Phys.\ Rev.}\ D {\bf 87} 014004.

\bibitem{AguilarSaavedra:2011ck}
  Aguilar-Saavedra J A and Santiago J 2012
  {\it Phys.\ Rev.}\ D {\bf 85} 034021.

\bibitem{Bai:2011ed}
  Bai Y, Hewett J L, Kaplan J and Rizzo T G 2011
  {\it JHEP} {\bf 1103} 003.

\bibitem{Ipek:2013zi}
  Ipek S 2013
  {\it Preprint} arXiv:1301.3990 [hep-ph].

\bibitem{Gresham:2012kv}
  Gresham M I, Shelton J and Zurek K M 2012
  {\it Preprint} arXiv:1212.1718 [hep-ph].

\bibitem{Blum:2011fa}
  Blum J Hochberg Y and Nir Y 2011
  {\it JHEP} {\bf 1110} 124.

\bibitem{Gresham:2012wc}
  Gresham M I, Kim I W, Tulin S and Zurek K M 2012
  {\it Phys.\ Rev.}\ D {\bf 86} 034029.

\bibitem{Skands:2012mm}
  Skands P, Webber B and Winter J 2012
  {\it JHEP} {\bf 1207} 151.

\bibitem{Brodsky:2012sz}
  Brodsky S J and Wu X G 2012
  {\it Phys.\ Rev.}\ D {\bf 86} 014021.

\bibitem{SH}
Doyle A C 1890
{\it The Sign of Four} (London: Spencer Blackett)

\end{thebibliography}
\end{document}